\documentclass[fleqn,usenatbib,useAMS]{mnras}

\usepackage{newtxtext,newtxmath}

\usepackage[T1]{fontenc}
\usepackage{ae,aecompl}

\usepackage[utf8]{inputenc}

\usepackage{graphicx}	
\usepackage{amsmath}	

\usepackage{multirow}
\usepackage{multicol}
\usepackage{booktabs}
\usepackage{siunitx}
\usepackage{xcolor}
\usepackage{hhline}

\usepackage{textcase}
\usepackage{braket}

\usepackage[caption=false]{subfig}

\usepackage{multirow}
\usepackage{caption}
\usepackage{adjustbox}

\newcommand\Tstrut{\rule{0pt}{2.6ex}}       
\newcommand\Bstrut{\rule[-1.2ex]{0pt}{0pt}} 

\newcommand{\Mpc}{\mathrm{Mpc}}

\newcommand\ee{\end{equation}}
\newcommand\be{\begin{equation}}
\newcommand\eea{\end{eqnarray}}
\newcommand\bea{\begin{eqnarray}}
\newcommand{\bV}{\mathbf{V}}

\newcommand{\bn}{\mathbf{n}}
\newcommand{\bx}{\mathbf{x}}

\newcommand{\bk}{\mathbf{k}}  

\newcommand{\B}{\textrm{B}}
\newcommand{\F}{\textrm{F}}
\newcommand{\HH}{\mathcal{H}}
\newcommand{\fe}{f^{\rm evo}}

\title[The relativistic dipole of a galaxy cross-correlation with DESI]{A case study for measuring the relativistic dipole of a galaxy cross-correlation with the Dark Energy Spectroscopic Instrument}

\author[Bonvin, Lepori, Schulz, Tutusaus, Adamek \& Fosalba]{Camille Bonvin,$^{1}$ Francesca Lepori,$^{2}$ Sebastian Schulz,$^{2}$ Isaac Tutusaus,$^{3,1}$ Julian Adamek,$^{2}$\newauthor and Pablo Fosalba$^{4,5}$
\\
$^{1}$D\'epartement de Physique Th\'eorique and Center for Astroparticle Physics (CAP), University of Geneva, 24 quai Ernest Ansermet, 1211 Gen\`eve, Switzerland\\
$^{2}$Institute for Computational Science, Universit\"at Z\"urich, Winterthurerstrasse 190, 8057 Z\"urich, Switzerland\\
$^{3}$Institut de Recherche en Astrophysique et Plan\'etologie (IRAP), Universit\'e de Toulouse, CNRS, UPS, CNES, 14 Av. Edouard Belin, 31400 Toulouse, France\\
$^{4}$Institute of Space Sciences (ICE, CSIC), Campus UAB, Carrer de Can Magrans, s/n, 08193 Barcelona, Spain\\
$^{5}$Institut d'Estudis Espacials de Catalunya (IEEC), Carrer Gran Capit\`a 2-4, 08193 Barcelona, Spain
}

\date{Accepted XXX. Received YYY; in original form ZZZ}

\pubyear{2023}

\begin{document}
\label{firstpage}
\pagerange{\pageref{firstpage}--\pageref{lastpage}}
\maketitle
\begin{abstract}
The data on spectroscopic galaxy clustering collected by the Dark Energy Spectroscopic Instrument (DESI) will allow the significant detection of subtle features in the galaxy two-point correlation in redshift space, beyond the ``standard'' redshift-space distortions. Here we present an independent assessment of the detectability of the relativistic dipole in the cross-correlation of two populations of galaxies if they would be selected from the Bright Galaxy Survey (BGS) of DESI. We build synthetic galaxy catalogues with the characteristics of the BGS using the light cone of a relativistic $N$-body simulation. Exploring different ways of splitting the populations of galaxies we find that with an unequal split with more bright galaxies than faint galaxies the detectability is significantly boosted, reaching 19\,$\sigma$ in the redshift bin $0.2 \lesssim z \lesssim 0.3$ and expected to be even higher at lower redshift. Moreover, we find that the measured dipole agrees very well with the prediction of relativistic effects from linear theory down to separations of $\sim$ 30\,Mpc/$h$.
\end{abstract}


\section{Introduction}

One of the main goals of galaxy surveys is to test the theory of gravity and to determine if the accelerated expansion of the Universe is due to a dark energy component or to deviations from General Relativity (GR) at cosmological scales. Spectroscopic redshift surveys have already delivered invaluable information on this question by providing very precise measurements of the galaxy peculiar velocities through the so-called redshift-space distortions (RSD)~\citep{Kaiser:1987qv, Beutler:2012px, Blake:2013nif,Okumura:2015lvp, BOSS:2016ntk, BOSS:2016psr, BOSS:2016off, BOSS:2016teh,Pezzotta:2016gbo,eBOSS:2020yzd}. 
These observations have allowed us to measure the growth rate of structure, which is a key quantity to test the theory of gravity, and has so far corroborated the validity of GR. The coming generation of surveys is expected to drastically improve on these measurements by covering wider areas of the sky, enlarging the redshift range of observation, and increasing the number density of galaxies~\citep{DESI:2016fyo, EUCLID:2011zbd, Amendola:2016saw, Maartens:2015mra}. 

Importantly, these future surveys also have the potential to measure novel effects in galaxy clustering that are too small to be detected at cosmological scales by current surveys. Over the past ten years it has indeed been shown that various distortions to the galaxy number counts, besides RSD, contribute to the observed clustering of galaxies~\citep{Yoo:2009au, Yoo:2010ni, Bonvin:2011bg, Challinor:2011bk,Jeong:2011as}. These distortions, called relativistic effects,\footnote{As discussed later, some of the so-called relativistic effects arise uniquely from GR, others from special relativity. Some of them can also be recovered from the classical Doppler effect in a Newtonian derivation.} impact our maps of the sky by modifying the redshift and angular position of galaxies. Fortunately, these distortions have a negligible impact on the monopole, quadrupole and hexadecapole of the correlation function (or power spectrum), and therefore do not contaminate the measurement of the growth rate of structure\footnote{Note that the contribution from lensing magnification does impact the measurement of the growth rate at high redshift~\citep{Jelic-Cizmek:2020pkh}. This effect can however be included in the modelling as a fixed contamination (independent on cosmology), which is enough to de-bias the measurement of the growth rate.} \citep{Tansella:2017rpi}.
Interestingly however, some of these distortions have the property to generate odd multipoles in the cross-correlation of two populations of galaxies, providing a robust way to isolate them from RSD and to  measure them, see \citet{McDonald:2009ud, Bonvin:2013ogt}. The detectability of these odd multipoles has been studied using linear perturbation theory, showing that the dipole should be measurable with a large signal-to-noise with the coming generation of surveys like the Dark Energy Spectroscopic Instrument\footnote{\url{https://www.desi.lbl.gov}} (DESI) and the Square Kilometer Array \footnote{\url{https://www.skao.int}}~\citep{Bonvin:2015kuc,Hall:2016bmm, Lepori:2017twd, Bonvin:2018ckp, Lepori:2019cqp}. 

In the non-linear regime, the relativistic dipole has been detected in BOSS with a significance of 2.8\,$\sigma$ at scales of $\sim$ 10\,Mpc/$h$~\citep{Alam:2017izi}. It has also been measured in numerical simulations for the cross-correlation of two populations of halos, see~\cite{Breton:2018wzk, Beutler:2020evf}. In addition, relativistic effects have also been detected in clusters with a significance of 3\,$\sigma$~\citep{Wojtak:2011ia}. This significance however decreases once systematic effects are taken into account~\citep{Sadeh:2014rya}.

The aim of this paper is first to provide a robust and realistic statement on the detectability of the dipole in the linear regime with the Bright Galaxy Survey (BGS) of DESI, by 
constructing and analysing a synthetic
catalogue with the characteristics of the BGS. This allows us to show that the dipole will be detectable with high significance in the BGS, namely 19\,$\sigma$ at $z=0.25$ and 8\,$\sigma$ at $z=0.35$. We also compare the dipole measured in our simulation with the theoretical linear prediction and we find a very good agreement for separations above $30$\,Mpc/$h$, 
validating the use of linear theory at those scales. Furthermore, we explore various ways of splitting the populations of galaxies to boost the significance of the dipole. We show that the optimal situation at low redshift is to split the catalogue into 90 percent of bright galaxies, and 10 percent of faint galaxies. This unequal split allows us to boost the significance by 30 percent at $z=0.25$. Our analysis shows therefore that the BGS should provide the very first detection of the relativistic dipole in the linear regime.

The rest of the paper is organised as follows. In Section~\ref{sec:theory}, we show the different relativistic terms that contribute to the clustering of galaxies and to the dipole. We also provide a theoretical expression for the covariance of the dipole. In Section~\ref{sec:catalog} we build the synthetic BGS catalogue from the light cone of an $N$-body simulation that was performed with the relativistic code \textit{gevolution}. To include all relevant effects, we assign luminosities to halos and use the relativistic luminosity distance to translate these into observed fluxes. In Section~\ref{sec:split} we explore different ways of splitting  the populations of galaxies and we present a method to measure the magnification bias and evolution bias of these samples that are needed for the theoretical prediction of the dipole. In Section~\ref{sec:measurement} we show the dipole and covariance measured in the catalogues and compare it with the theoretical predictions. Finally, we compute the significance of the dipole for different splits between bright and faint galaxies. We  conclude in Section~\ref{sec:conclusion}. 

\section{Theoretical model}
\label{sec:theory}

\subsection{The relativistic dipole}

Galaxy surveys provide a measurement of the fluctuations of the galaxy number counts
\be
\Delta\equiv\frac{N(\bn,z)-\bar N(z)}{\bar N(z)}\, ,
\ee
where $N$ denotes the number of galaxies per pixel in direction $\bn$ and at redshift $z$, and $\bar N$ is the average number of galaxies per pixel at redshift $z$. The full theoretical modelling for the observable $\Delta$ in the linear regime has been derived by various groups~\citep{Bonvin:2011bg,Challinor:2011bk,Yoo:2009au,Jeong:2011as}, who showed that this quantity is not only affected by density fluctuations and redshift-space distortions (RSD), but also by a number of other distortions generated by galaxy peculiar velocities and by the two gravitational potentials. More precisely, we have
\begin{align}
&\Delta(\mathbf{n},z)=b\cdot \delta-\frac{1}{\mathcal{H}}\partial_r(\mathbf{V}\cdot\mathbf{n})
+\left(5s-2\right)\int_0^{r}\!\!dr' \frac{r-r'}{2rr'}\Delta_\Omega(\Phi+\Psi)\nonumber\\
&+\left(1-\frac{\dot{\mathcal{H}}}{\mathcal{H}^2}+\frac{5s-2}{r\mathcal{H}} -5s+f^{\rm evo}\right)\mathbf{V}\cdot\mathbf{n}+\frac{1}{\mathcal{H}}\dot{\mathbf{V}}\cdot\mathbf{n}+\frac{1}{\mathcal{H}}\partial_r\Psi\nonumber\\
&+\frac{2-5s}{r}\int_0^{r}\!\!dr'(\Phi+\Psi)+\left(3 - f^{\rm evo}\right)\mathcal{H}\,\nabla^{-2}(\nabla\cdot\mathbf{V})+(5s-2)\Phi\nonumber\\
&+\frac{1}{\mathcal{H}}\dot{\Phi}+\Psi+\left(\frac{\dot{\mathcal{H}}}{\mathcal{H}^2}+\frac{2-5s}{r\mathcal{H}}+5s-f^{\rm evo} \right)\left[\Psi+\int_0^{r}\!\!dr'(\dot{\Phi}+\dot{\Psi})\right]\,.\label{eq:Delta}
\end{align}
Here $r=r(z)$ is the comoving distance to redshift $z$, a dot denotes derivative with respect to conformal time $\eta$, $\HH$ is the Hubble parameter in conformal time and $\Delta_\Omega$ is the angular Laplacian. Eq.~\eqref{eq:Delta} is written in terms of gauge-invariant quantities, namely the two Bardeen potentials\footnote{These Bardeen potentials reduce to the spatial and temporal metric perturbations in Newtonian gauge: $\mathrm ds^2 = -a^2(\eta)\big[1+2 \Psi(\bx,\eta)\big]d \eta^2 + a^2(\eta)\big[1- 2 \Phi(\bx,\eta)\big]d \mathbf{x}^2$.} $\Phi$ and $\Psi$, the gauge-invariant density fluctuation $\delta$ (corresponding to the ordinary density perturbation in the comoving gauge), and the gauge-invariant peculiar velocity $\bV$ (corresponding to the ordinary peculiar velocity in the Newtonian gauge).\footnote{We work here with units such that $c=1$}
The functions $b(z)$, $s(z)$  and $f^{\rm evo}(z)$  are the linear galaxy bias, the magnification bias and the evolution bias of the population under consideration.

The first two terms in the first line of Eq.~\eqref{eq:Delta} are the standard density and RSD contributions, that give rise to a monopole, quadrupole and hexadecapole in the two-point correlation function (or power spectrum). The third term is the contribution from lensing magnification. This contribution has been measured in the distribution of quasars~\citep{Scranton:2005ci} and galaxies at high redshift~\citep{Liu:2021gbm}. It is subdominant in current spectroscopic galaxy surveys, but will become relevant for the coming generation of surveys at high redshift~\citep{Jelic-Cizmek:2020pkh}. Since we focus here on the BGS sample at $z\leq 0.5$, lensing magnification is completely negligible~\citep{Bonvin:2013ogt}. 
The last three lines are the so-called relativistic effects. Note that these terms are not relativistic in the sense that they would involve higher powers of $V$. 
They account for both the impact of spatial distortion, $\Phi$, and time distortion, $\Psi$, on the light's energy, as well as several light-cone projection effects related to peculiar motion along the line of sight.
Among the relativistic terms, the ones in the second line are the dominant contributions. They are indeed suppressed by one power $\HH/k$ with respect to density and RSD, whereas the last two lines are suppressed by two powers, $(\HH/k)^2$. It was shown that the terms in the second line do not significantly contribute to the standard monopole, quadrupole and hexadecapole~\citep{Tansella:2017rpi}. However, they have the special property to generate a dipole in the correlation function, allowing us to isolate them from the dominant density and RSD. The first two terms in the second line are Doppler contributions, whereas the last term is the contribution from gravitational redshift, which depends directly on $\Psi$.

To measure the relativistic dipole, it is necessary to correlate two populations of galaxies. Here we split the sample into a bright population (B), containing all galaxies brighter than a chosen flux threshold, and a faint population (F). The dipole in the correlation function then reads
\begin{align} \label{eq:dipole_gen}
&\xi_1(d,z) = -(b_\B - b_\F) \int \frac{dk k^2}{2 \upi^2}\frac{k}{\HH} P_{\delta \Psi} (k, z)j_1 (kd)\\
&+\int \frac{dk k^2}{2 \upi^2}\Bigg[(b_\B \alpha_\F - b_\F \alpha_\B) P_{\delta V} (k, z)  \nonumber\\
&  +(b_\B - b_\F)\frac{1}{\HH}P_{\delta \dot{V}} (k,z)  + \frac 35 (\alpha_\B - \alpha_\F) \left(\frac{k}{\mathcal{H}}\right) P_{V V} (k,z)\Bigg] j_1 (kd) \nonumber \\
& + \frac{2}{5}\frac dr (b_\B - b_\F)  \int \frac{dk k^2}{2 \upi^2} \frac{k}{\HH}  P_{\delta V} (k, z) j_2 (kd)\, , \nonumber 
\end{align}
where $d$ denotes the separation between galaxies and  
\begin{equation}
    \alpha_{\rm L} \equiv 1 - 5 s_{\rm L} + \frac{5 s_{\rm L} -2}{r \mathcal{H}} - \frac{\dot{\mathcal{H}}}{\mathcal{H}^2} + f_{\rm L}^\mathrm{evo}\, ,\quad\mbox{for} \quad {\rm L}=\B,\F\, .
\end{equation}

The first line of Eq.~\eqref{eq:dipole_gen} contains the gravitational redshift contribution, which depends on the correlation between $\Psi$ and $\delta$. This term is due to the fact that the light emitted by galaxies is shifted to the red by the large-scale distortion of time $\Psi$. The second and third lines contain Doppler effects, proportional to the galaxy peculiar velocity $V$. The peculiar velocity contributes in different ways to the dipole. First, it shifts the observed redshift of galaxies, displacing their radial positions on the map. Since the number density of galaxies evolves with redshift, this impacts the observed clustering. There are two types of evolution that enter into Eq.~\eqref{eq:dipole_gen}: first the  dilution of galaxies due to the expansion in the Universe, which results in terms proportional to $\HH$ and $\dot\HH$, and second the physical merging or creation of galaxies, which is encoded in the function
\begin{align}
\fe_{\rm L}= \left.\frac{\partial \mathrm{ln} \bar{n}_{\rm L}}{\partial \mathrm{ln} a}\right|_{\bar{L}_\mathrm{cut}(z)},   \label{eq:fevo-def}
\end{align}
where $\bar{n}_{\rm L}$ is the mean comoving number density of objects and $\bar{L}_\mathrm{cut}(z)$ is the luminosity cut that defines the population ${\rm L}$. 
In the literature, $\fe_{\rm L}$ is referred to as `evolution bias'.
In addition, the peculiar velocity also contributes through its impact on the observed flux of galaxies. As an example, suppose that there are two galaxies at the same redshift $z$ and with the same intrinsic luminosity: galaxy 1 has a peculiar velocity towards us, whereas galaxy 2 has no peculiar velocity. This means that galaxy 1 is physically further away from us than galaxy 2. As a consequence its observed flux will be smaller than the one of galaxy 2. Since surveys are flux limited, this implies that galaxy 1 may not be detectable whereas galaxy 2 is, even though they both have the same intrinsic luminosity and the same redshift. The amplitude of this effect is governed by the magnification bias $s_{\rm L}$ of the population L.
Finally, the last line of Eq.~\eqref{eq:dipole_gen} contains the wide-angle contribution, due to RSD. Note that this contribution is suppressed by a factor $d/r$ with respect to the RSD contribution in the even multipoles. Since $d/r\sim \HH/k$, this contribution is of the same order of magnitude as the relativistic effects and therefore needs to be properly accounted for in the dipole. 

Eq.~\eqref{eq:dipole_gen} is valid for any theory of gravity and for any model of dark matter. It only assumes that photons propagate on null geodesics of a perturbed Friedman metric. In models where the continuity equation for baryonic and dark matter is valid (i.e.\ there is no exchange of energy between matter and another component) we have
\begin{align}
V(\bk, z)=-\frac{\HH(z)}{k}f(z)\delta(\bk, z)\, , \quad \mbox{with}\quad f=\frac{d\ln \delta}{d\ln a}
\end{align}
the growth rate. If in addition matter obeys Euler's equation (i.e.\ there is no momentum exchange nor fifth force acting on dark matter) we have
\begin{align}
\dot{V}(\bk, z)+\HH(z) V(\bk, z)-k\Psi(\bk, z)= 0\,, \label{eq:euler}
\end{align}
and the dipole reduces to
\begin{align}
&\xi_1(d,z) = \frac{\HH}{\HH_0} 
\Bigg[(b_\B-b_\F) f \left(\frac{2}{r\HH} + \frac{\dot{\HH}}{\HH^2}\right)
+ 3 (s_\F-s_\B) f^2 \left( 1-\frac{1}{r\HH}\right)\nonumber\\
&+ 5 (b_\B s_\F-b_\F s_\B) f \left( 1-\frac{1}{r\HH}\right)
+\frac{3}{5}\left(\fe_\B-\fe_\F\right)f^2\nonumber\\
&+\left(b_\F\fe_\B-b_\B\fe_\F\right)f\Bigg] \nu_1(d,z)-\frac{2}{5} (b_\B-b_\F) f \frac{d}{r}  \mu_2(d,z) \, ,\label{eq:dip}
\end{align}
where
\begin{align}
&\nu_1(d,z)=\frac{1}{2\upi^2}\int d k k\HH_0 P_{\delta\delta}(k,z)j_1(kd)\, , \label{eq:nu1} \\
&\mu_2(d,z)=\frac{1}{2\upi^2}\int d k k^2 P_{\delta\delta}(k,z)j_2(kd)\, . \label{eq:mu2}
\end{align}
The dipole in Eq.~\eqref{eq:dip} depends on the characteristics of the two populations of bright and faint galaxies, more precisely on the biases, magnification biases and evolution biases of these populations. In the next section we describe how these parameters are modelled and measured in our catalogues.

\subsection{Magnification bias and evolution bias}
\label{sec:biases-th}
For one population of galaxies, the magnification bias is simply defined as the slope of the cumulative number of galaxies, evaluated at the flux limit of the survey $F_{\rm lim}$:
\begin{align}
s(z,F_{\rm lim})\equiv -\frac{2}{5}\frac{\partial \ln \bar{N}(z,F\geq F_{\rm lim})}{\partial\ln F_{\rm lim}}\, .
\label{eq:sbias-def}
\end{align}
For two populations of galaxies, the situation is different since there are two flux limits: the flux limit of the survey and the flux threshold used to split the population into bright and faint galaxies $F_{\rm cut}$. 

The magnification bias of the bright population is simply given by
\begin{align}
s_\B(z)=s(z,F_{\rm cut})= -\frac{2}{5}\frac{\partial \ln \bar{N}(z,F\geq F_{\rm cut})}{\partial\ln F_{\rm cut}}\, ,
\end{align}
where $\bar{N}(z,F\geq F_{\rm cut})=\bar{N}_\B(z)$ is the mean number of bright galaxies per pixel. 
To derive the magnification bias of the faint population, we calculate the variation of the number of faint galaxies due to luminosity perturbations at both flux limits. The number of faint galaxies per pixel is given by
\begin{align}
\label{eq:NF}
 N_\F(z,\bn)&\equiv N(z,\bn,F_{\rm cut}> F\geq F_{\rm lim})\,.
 \end{align}
The fluxes, $F_{\rm cut}$ and $F_{\rm lim}$, correspond to different intrinsic luminosity thresholds in different directions: $L_{\rm cut}(z,\bn)=\bar{L}_{\rm cut}(z)+\delta L_{\rm cut}(z,\bn)$ and $L_{\rm lim}(z,\bn)=\bar{L}_{\rm lim}(z)+\delta L_{\rm lim}(z,\bn)$. Inserting this into~\eqref{eq:NF} and expanding around the background luminosity thresholds we obtain
\begin{align}
 N_\F(z,\bn)&=N(z,\bn,L\geq \bar{L}_{\rm lim})-\frac{5}{2}s(z, F_{\rm lim})\bar{N}(z) \frac{\delta L_{\rm lim}}{\bar{L}_{\rm lim}}(z,\bn)\nonumber\\
 &-N(z,\bn,L\geq \bar{L}_{\rm cut})+\frac{5}{2}s(z, F_{\rm cut})\bar{N}_\B(z) \frac{\delta L_{\rm cut}}{\bar{L}_{\rm cut}}(z,\bn)\, ,\nonumber
\end{align}
where $\bar{N}$ is the mean number of galaxies (bright and faint) per pixel. Using that the fractional fluctuations in luminosity are independent of $L$ (they are proportional to the fluctuations of the luminosity distance), we obtain for the magnification bias of the faint population
\begin{align}
s_\F(z)&\equiv s(z,F_{\rm lim})\frac{\bar{N}(z)}{\bar{N}_\F(z)}-s(z,F_{\rm cut})\frac{\bar{N}_\B(z)}{\bar{N}_\F(z)}\,.
\label{eq:sfaint}
\end{align}
We see that this magnification bias depends not only on the flux threshold $F_{\rm cut}$ and on the flux limit of the survey $F_{\rm lim}$, but also on the fraction of bright and faint galaxies. This dependence can be used to boost the amplitude of the dipole. Note that in the case where there is no flux limit and $\bar{N}_\B=\bar{N}_\F$ we obtain from Eq.~\eqref{eq:sfaint} that $s_\F=-s_\B$. This simply reflects the fact that Doppler effects have an opposite impact on the bright and faint samples: they remove galaxies with a flux below $F_{\rm cut}$ from the bright sample and move them to the faint sample (or vice-versa).

The amplitude of the dipole is also affected by the evolution bias,
which models the departure from the conservation of the number of sources
in a comoving volume, see Eq.~\eqref{eq:fevo-def}. Since we want to express $\fe_{\rm L}$ in terms of observable quantities, we rewrite the mean comoving number density of sources in terms of the average number of sources  
per pixel in the redshift bin $\bar{N}_{\rm L}$.
Neglecting perturbations, $\bar{N}_{\rm L}$ and
$n_{\rm L}$ are simply related by a geometrical factor, i.e. $\bar{n}_{\rm L}  \propto \bar{N}_{\rm L} \frac{(1+z)\HH}{r^2}$, which leads to
\begin{align}
\fe_{\rm L}= - \left.\frac{\partial \mathrm{ln}  \bar{N}_{\rm L}}{\partial \mathrm{ln}(1+z)} \right|_{\bar{L}_\mathrm{cut}(z)}
 + \frac{\dot\HH}{\HH^2} + \frac{2}{r\HH} - 1. 
\label{eq:fevo-2}
\end{align}
The derivative in the first term of Eq.~\eqref{eq:fevo-2} has to be evaluated at the luminosity cut $\bar{L}_{\rm cut}(z)$. However, we do not observe the intrinsic luminosity of galaxies, we rather observe their fluxes. Intrinsic luminosity and observed flux are related by $\bar{L}_{\rm cut}(z) = 4\upi \bar{d}^2_{\rm L}(z) F_\mathrm{cut}$, where $\bar{d}_{\rm L}(z) = (1+z)\,r(z)$ is the background luminosity distance (since $f^{\rm evo}$ multiplies first order quantities)
and $F_\mathrm{cut}$ is the flux cut of our catalogue. To evaluate Eq.~\eqref{eq:fevo-2} at $F_{\rm cut}$ we need to account for the fact that the luminosity cut is redshift-dependent. We 
therefore transform the partial derivative over redshift into a total derivative, that we can then evaluate at a fixed flux cut $F_{\rm cut}$: 
\begin{align}
\left.\frac{\partial \mathrm{ln}  \bar{N}_{\rm L}}{\partial \mathrm{ln}(1+z)} \right|_{\bar{L}_\mathrm{cut}(z)}&=  \frac{d \mathrm{ln}  \bar{N}_{\rm L}(z,\bar{L}_{\rm cut}(z))}{d \mathrm{ln}(1+z)} -
\frac{\partial \mathrm{ln}  \bar{N}_{\rm L}}{\partial \mathrm{ln}\bar{L}_{\rm cut}} \frac{d\ln\bar{L}_{\rm cut}}{d\ln(1+z)}\nonumber\\
&=\frac{d \mathrm{ln}  \bar{N}_{\rm L}(z,F_\mathrm{cut})}{d \mathrm{ln}(1+z)} +\left(1+\frac{1}{r\HH} \right)5s_{\rm L}\, .
\label{eq:fevobis}
\end{align}
Inserting this into~\eqref{eq:fevo-2} we obtain
\begin{align}
\fe_{\rm L}= - \frac{d \mathrm{ln}  \bar{N}_{\rm L}(z,F_\mathrm{cut})}{d \mathrm{ln}(1+z)}
 + \frac{\dot\HH}{\HH^2} + \frac{2-5 s_{\rm L}}{r\HH} - 5 s_{\rm L} - 1,
\label{eq:fevo-3}
\end{align}
which is in agreement with Eq.~(2.26) of~\cite{Maartens:2021dqy}.

\subsection{Theoretical covariance}
\label{sec:theory_cov}

The theoretical covariance of the dipole has been derived in~\cite{Hall:2016bmm}. It contains a shot-noise contribution and a mixed contribution, due to the product of shot noise and cosmic variance. The pure cosmic-variance contribution from density and RSD vanishes, due to symmetry. However, the pure cosmic variance from relativistic effects does not vanish. This term is usually neglected since it is subdominant compared to the mixed term. However, we have checked that for the lowest redshift bin of DESI, this contribution is relevant at large separations. We therefore include it in our forecasts. Following the derivation presented in~\cite{Hall:2016bmm}, we find for the covariance
\begin{align}
&{\rm cov}^{\rm th}_{ij}\equiv{\rm cov}^{\rm th}\big[\xi_1(d_i,z),\xi_1(d_j,z)\big]=
\frac{3 V}{4\upi N^{\rm tot}_\B N^{\rm tot}_\F \ell_{p} d_i^2}\delta_{ij}\label{eq:cov_theo}\\
&+\frac{9}{2}\!\left[\frac{1}{N^{\rm tot}_\F}\!\left(\frac{b_\B^2}{3}+\frac{2b_\B f}{5}+\frac{f^2}{7} \right)+\frac{1}{N^{\rm tot}_\B}\!\left(\frac{b_\F^2}{3}+\frac{2b_\F f}{5}+\frac{f^2}{7} \right)\right]G_{ij} \nonumber\\ 
&+\frac{9}{4V}\left(\frac{\HH}{\HH_0}\right)^2\Bigg[\frac{2}{5} (b_\F c_\B - b_\B c_\F)^2 + \frac{4}{7} (c_\B - c_\F) (b_\F c_\B - b_\B c_\F) f\nonumber\\
&+ \frac{2}{9} (c_\B - c_\F)^2 f^2\Bigg]J_{ij}\, ,\nonumber
\end{align}
where $V$ is the volume of the redshift bin, ${N}^{\rm tot}_\B$ and ${N}^{\rm tot}_\F$ denote the total number of bright and faint galaxies in that bin and $\ell_p$ represents the pixel size (or the step over which separations are averaged). The functions $G_{ij}$ and $J_{ij}$ depend on the density power spectrum and are given by
\begin{align}
G_{ij}=&\frac{1}{\upi^2}\int d k \; k^2 P_{\delta\delta}(k,z)j_1(kd_i)j_1(kd_j)\, ,\\
J_{ij}=&\frac{1}{\upi^2}\int d k \; \HH_0^2 P_{\delta\delta}^2(k,z)j_1(kd_i)j_1(kd_j)\, ,
\end{align}
and 
\be
c_\mathrm{L}=\left(5 s_\mathrm{L}+ \frac{2 - 5 s_\mathrm{L}}{r \HH} +\frac{\dot\HH}{\HH^2}-f_{\rm L}^\mathrm{evo}\right) f\, ,\quad\mbox{for}\quad \mathrm{L}=\B,\F\,.
\ee
The first line of Eq.~\eqref{eq:cov_theo} contains the shot-noise contribution, the second line the mixed contribution from shot noise and cosmic variance, and the last two lines contain the cosmic variance from relativistic effects. The covariance between different separations $d_i\neq d_j$ is included in Eq.~\eqref{eq:cov_theo}, however the covariance between different redshift bins is neglected, as is usually the case in RSD analyses since the bins are wide enough and do not overlap. 

In Fig.~\ref{fig:covariance} we plot the different contributions to the error, for the specifications shown in Table~\ref{tab:spec} (case 1).  We see that the mixed contribution is important at all redshifts. At $z=0.25$ the cosmic variance from relativistic effect is relevant at large scales. It would even become more important in the two lowest bins of DESI, at $z=0.05$ and $z=0.15$ that we are not considering here. This shows that to properly estimate the covariance, one needs to include relativistic effects {when constructing catalogues from simulations. At $z=0.35$ and $z=0.45$ the relativistic cosmic variance is completely negligible. At $z=0.45$ the pure shot noise contribution becomes of the same order as the mixed contribution. From Fig.~\ref{fig:covariance}, we see therefore that contrary to the even multipoles, that will be cosmic variance limited in the BGS, the dipole is strongly affected by shot noise at all redshifts. Galaxy samples with high number densities are therefore crucial to optimally measure the dipole.

\begin{figure}
    \centering
    \includegraphics[width=0.47\textwidth]{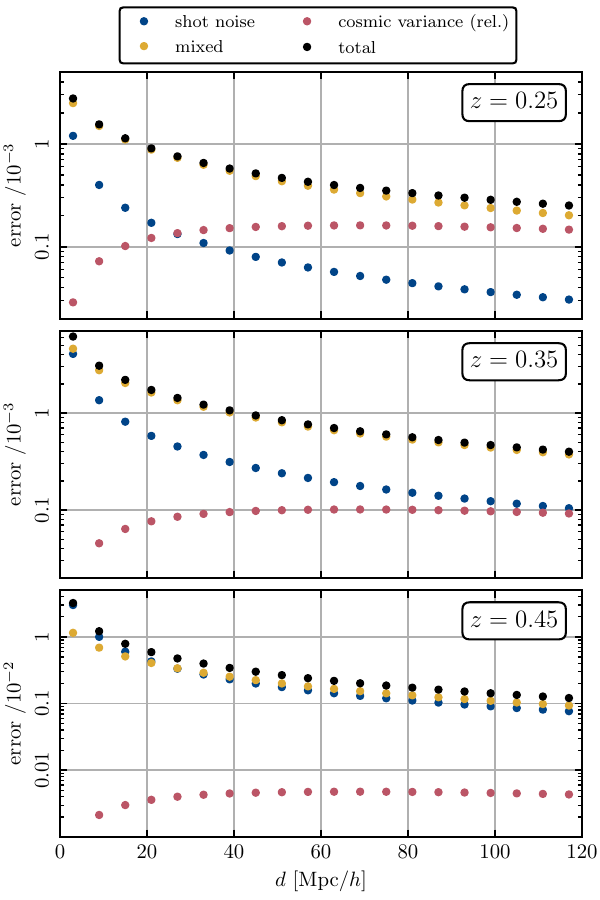}
    \caption{Theoretical error, $\sqrt{\mathrm{cov}^{\rm th}(d_i,d_i)}$, on the dipole at $z=0.25$ (top panel), $z=0.35$ (middle panel) and $z=0.45$ (bottom panel) for pixel size $\ell_p=6$\,Mpc/$h$. The mixed contribution is shown in yellow, the shot noise in blue, the cosmic variance from relativistic effects in pink and the total in black.}
    \label{fig:covariance}
\end{figure}

\section{Catalogues}
\label{sec:catalog}

\subsection{$N$-body simulations from {\it gevolution}}

In order to obtain a synthetic source catalogue that reproduces the relevant characteristics of the Bright Galaxy Survey of DESI, we use existing simulation data from the ``unity2'' run which was done with the code \textit{gevolution} \citep{Adamek:2015eda}. This $N$-body simulation of $5760^3$ particles in a volume of $(4032\,\Mpc/h)^3$ had a fixed spatial resolution of $700\,\mathrm{kpc}/h$ which is sufficient for having many galactic halos form but not enough to resolve their internal structure well. However, for the purpose of this work we are interested in two-halo correlations at scales at least an order of magnitude larger than the resolution scale. The ``unity2'' run provides a halo catalogue on a full-sky light cone up to redshift $z \simeq 0.8$ without any replications. The cosmology is partially based on a best fit to $\Lambda$CDM from \citet{Planck:2013pxb}, assuming a minimal-mass scenario for neutrinos, $\sum m_\nu \simeq 0.06\,\mathrm{eV}$. The parameters were $A_\mathrm{s} = 2.215\times 10^{-9}$ at the pivot scale $k_\mathrm{p} = 0.05\,\Mpc^{-1}$, $n_\mathrm{s} = 0.9619$, $h = 0.67$, $\Omega_\mathrm{b} = 0.049$ and $\Omega_\mathrm{m} = 0.319$.

The halo catalogue is constructed using the \textit{ROCKSTAR} halo finder \citep{Behroozi:2011ju} and we keep any identified subhalos to allow particularly large halos to be populated by more than one galaxy. The catalogue is post-processed using the ray tracer described in \citet{Lepori:2020ifz} to obtain the observed positions and redshifts for all objects. While the method takes into account even subleading relativistic effects, the most important contribution comes from the Doppler shift. The fact that our simulation pipeline uses GR at each stage helps maintaining internal consistency, but the Newtonian limit is also expected to work extremely well on the scales and redshifts of interest. This means that similar catalogues could be constructed from Newtonian simulations if there is a suitable method for generating data on the light cone. 

The left panel of Fig.~\ref{fig:cat-comoving} shows a small, $60\,\mathrm{Mpc}/h \times 60\,\mathrm{Mpc}/h$, patch of the light cone at redshift $z \simeq 0.35$ which corresponds roughly to a comoving distance of $1\,\mathrm{Gpc}/h$ from the observer. The thickness of the slice orthogonal to the figure plane is approximately $8\,\mathrm{Mpc}/h$. The region contains a void surrounded by some filaments and clusters. The right panel of Fig.~\ref{fig:cat-comoving} shows the same region in redshift space; the expansion of the void due to the Kaiser effect is clearly evident, as is the so-called Fingers-of-God effect that disperses the clusters along the line of sight due to nonlinear orbital motion.

\begin{figure*}
\centering
\includegraphics[width=\columnwidth,trim = 4mm 2mm 10mm 12mm, clip]{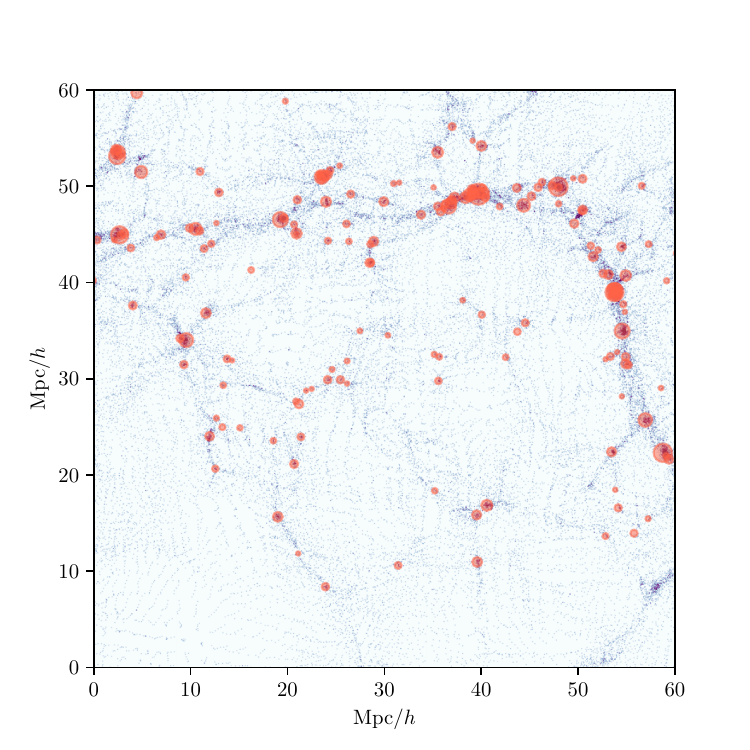}
\hfill
\includegraphics[width=\columnwidth,trim = 4mm 2mm 10mm 12mm, clip]{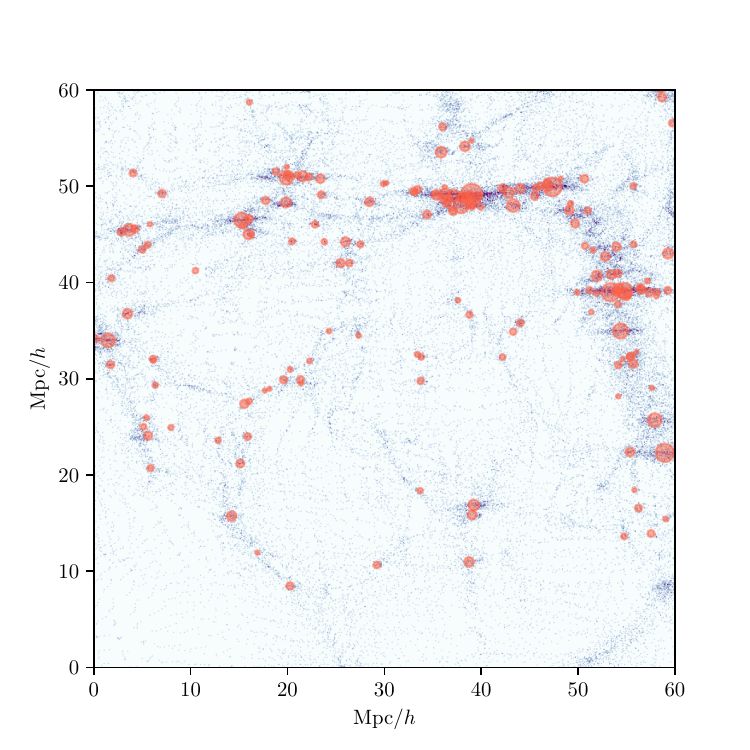}
\caption{Small patch of the light cone in comoving space (left panel) and redshift space (right panel) at redshift $z \simeq 0.35$. The density of particles is indicated in blue whereas halos are shown as red disks with an area proportional to their respective masses. The observer is located towards the left, at a distance of about $1\,\mathrm{Gpc}/h$.}
\label{fig:cat-comoving}
\end{figure*}

\subsection{Assigning luminosities and fluxes to halos}

Once the halo catalogues have been generated, the next step in the pipeline is to assign a luminosity to each one of the objects. We follow the approach presented in~\citet{Carretero:2014ltj} and use an abundance-matching technique with a luminosity function extracted from observations. In more detail, we first use the number of particles in every halo as a proxy for its mass. However, this is a discrete quantity, so we add a uniform noise between 0 and 1 to each number of particles. Secondly, using the halo mass function of our catalogue, we compute the cumulative mean number density of halos, $n_{\rm halo}(>M_{\rm min})$. On the other hand, we consider the best-fit of the modified Schechter function derived from the Sloan Digital Sky Survey \citep[SDSS,][]{SDSS:2002vxn,Blanton:2004zy}, and compute the mean number density of objects with luminosities brighter than a certain threshold,
\begin{equation}
    n_{\rm halo}(>L_{\rm t})=\int_{L_{\rm t}}^{\infty}\frac{\text{d}n}{\text{d}L}\,\text{d}L\,,
\end{equation}
where $L_{\rm t}$ stands for a luminosity threshold. Using an abundance-matching technique, we can extract a relation between the mass of a given halo, $M_{\rm halo}$ (or the noisy number of particles in our case), and the luminosity assigned to it, $L_{\rm halo}$. The relation is obtained by equalling both cumulative functions,
\begin{equation}
    n_{\rm halo}(>M_{\rm min}) = n_{\rm halo}(>L_{\rm t})\,.
\end{equation}

It is important to mention that the SDSS data considered to extract the luminosity function extends only to redshift $z\lesssim 0.1$. Therefore, we select a sphere around the observer with a radius of 300\,Mpc$/h$ to perform the abundance matching. When applying the relation between $M_{\rm halo}$ and $L_{\rm halo}$ to all the objects in the catalogue, some of these might be a bit less (or more) massive than the objects contained within the sphere of 300\,Mpc$/h$ radius. Because of that we use a constant extrapolation for less massive (fainter) and more massive (brighter) objects. However, we have checked that the amount of objects present in these range extensions is small.

We also note that this is a simpler approach than the one considered in~\citet{Carretero:2014ltj}, since we do not introduce any dispersion in our abundance matching. With an abundance-matching technique, the most luminous objects end up in the most massive halos. An additional scatter in the halo mass -- luminosity relation can be added to improve the agreement with clustering measurements of the brightest objects. Since we do not have data to compare with at this stage, we do not include such a dispersion in our catalogues. This may slightly impact the bias of the two populations that we will use in our forecast, and therefore the dipole signal. For example, adding scatter may shift some bright galaxies to the faint sample and vice versa. However, the dipole signal is dominated by pairs of galaxies with large bias difference (i.e.\ large luminosity difference), where we can expect that small random changes on the luminosity would not have a strong impact on average.

After assigning a luminosity to each one of the objects (halos) in the catalogue, we are interested in computing the associated fluxes: $F=L/(4\upi d_L^2)$. The luminosity distance is affected by inhomogeneities in the universe~\citep{Bonvin:2005ps}. To compute these perturbations in {\it gevolution}, we first relate the luminosity distance to the angular diameter distance $d_{\rm A}$
\begin{equation}
\label{eq:luminosity_distance}
    d_{\rm L} = (1+z)^2d_{\rm A}\,.
\end{equation}
The angular diameter distance at a given position is affected by gravitational lensing, Sachs Wolfe terms and time delay~\citep{Bernardeau:2009bm}. These terms can be computed exactly by ray tracing from the observer to the halos \citep{Lepori:2020ifz}. To obtain the luminosity distance at the position of the halos, we then simply multiply in Eq.~\eqref{eq:luminosity_distance} the angular diameter distance by the observed redshift (which is affected by Doppler effects and potential terms). 
Since below $z=0.5$, the only non-negligible contribution to the luminosity distance is the Doppler contribution, we can simplify the procedure by replacing the true angular diameter distance by the background one in Eq.~\eqref{eq:luminosity_distance}. This procedure neglects the contribution from gravitational lensing, which is anyway completely subdominant in the dipole at small redshift \citep[see e.g.\ Fig.~8 in][]{Bonvin:2013ogt}.

\subsection{Selection of sources to reproduce the DESI BGS}

The resulting catalogue is a list of objects, distributed over
the full sky, with associated observed coordinates, i.e.\ observed redshift and angular positions, true comoving positions, luminosities and observed fluxes. 
In order to build a catalogue that is adapted to the specifics of the DESI Bright Galaxy Survey, we split the catalogue in five redshift bins in the range
$z \in [0, 0.5]$, with equal width $\Delta z = 0.1$. In each redshift bin, we select
objects to i) match the number density of the BGS expected for DESI inside the bins, and ii) match the expected bias for DESI at the mean redshift of the bin.
We use the target specifics for the DESI BGS described in \citet{DESI:2016fyo}. 

\begin{table}
\caption{Number of sources in our complete catalogue, $N_\mathrm{presel}$, compared to the target specifics of the DESI BGS.
$\frac{dN_\mathrm{BGS}}{dz\,d\Omega}$ denotes the galaxy number density in redshift and solid angle, $N_\mathrm{target}$ represents the number of sources expected for a survey with the BGS number density distributed over the full sky, $N_\mathrm{BGS}$ 
is the expected number of sources in the DESI BGS, and $b_\text{BGS}$ is the forecast bias, estimated as $b_\mathrm{BGS}(z) = 1.34/D(z)$, where $D(z)$ is the linear growth factor normalised at redshift $z = 0$. 
}
\begin{center}
\adjustbox{max width=\columnwidth}{
\begin{tabular}{cc c c c c}
 \toprule
\multirow{2}*{$\bar{z}$} & $N_\mathrm{presel}$ & \multirow{2}*{$\!\!\!\!\!\!\frac{dN_\mathrm{BGS}}{dz\,d\Omega} [\mathrm{deg}^{-2}]\!\!\!\!\!\!$}  & $N_\mathrm{target}$  & $N_\mathrm{BGS}$ & \multirow{2}*{$b_\text{BGS}$}\\
& (full-sky)\Bstrut & & (full-sky) & \!($14000\,\mathrm{deg}^2$)\! & \\
 \midrule
 0.05  &    687'602\Tstrut & $1165$ & 4'805'974 &  1'631'000 & $1.38$  \\
 0.15  &  4'480'034 & $3074$ & 12'681'172 &  4'303'600 & $1.45$  \\
 0.25  & 11'213'239 & $1909$ & 7'875'197 & 2'672'600 & $1.53$  \\
 0.35  & 19'977'561 & $732$  & 3'019'719 & 1'024'800 & $1.61$  \\
 0.45  & 29'592'462 & $120$  &  495'036  & 168'000 & $1.70$  \\
\bottomrule
\end{tabular}}
\end{center}
\label{tab:target}
\end{table}

In Table~\ref{tab:target}, we report the number of sources
in our complete catalogue, for each redshift bin, compared to the 
number of sources expected in the DESI catalogue over 14'000 square degrees ($N_\mathrm{BGS}$)
and the number of sources that we want to include in our synthetic full-sky 
catalogue ($N_\mathrm{target}$). We represent the mean redshift of the bin as $\bar{z}$.
At low redshift, for the two bins centered at $\bar{z} = 0.05$ and $\bar{z} = 0.15$, 
the number of objects in our complete catalogue is smaller than the number of target sources.\footnote{This is due to the specifics of the simulation that we are using, namely to the limited resolution and the fact that we do not include satellite galaxies.} Therefore, in our analysis, we only consider the three redshift bins centred in $\bar{z} = 0.25, 0.35, 0.45$. The rationale for this is that if we can show that the dipole is detectable in these three redshift bins, adding the two lowest ones will only increase the significance of the detection.

In each of these bins, we assign to each object a probability to be selected
between $0$ and $1$, depending on their luminosity. 
We use as ansatz for the probability $P(\mathrm{log}L_i)$ a smoothed step function, inspired by Halo Occupation Distribution models,
\begin{equation}
P(\mathrm{log}L_i) = \frac{1}{2}\left[ 1 + \mathrm{erf}\left(\frac{\mathrm{log}L_i-\mathrm{log}L_\mathrm{min}}{\sigma_{L_\mathrm{min}}}\right)\right], \label{eq:hod-prob}
\end{equation}
where `$\mathrm{erf}$' denotes the error function, while $\mathrm{log}L_\mathrm{min}$ and $\sigma_{L_\mathrm{min}}$ are two 
parameters to be determined. They represent the logarithmic luminosity of sources that have $50\%$ probability to be selected and the smoothing of the step function, respectively. 
In order to estimate the values of these free parameters, we use an iterative method. In each redshift bin, we choose an initial guess for the value of $\sigma_{L_\mathrm{min}}$. We then use a root-finding method to determine the value of $\mathrm{log}L_\mathrm{min}$ such that
$\sum_i P(\mathrm{log}L_i) = N_\mathrm{target}$. 
This ensures that the number of sources selected from our probability function roughly matches our target $N_\mathrm{target}$. 
For these initial guesses, we select our sample according
to $P(\mathrm{log}L_i)$, and we estimate the galaxy bias of the sample as described in Sec.~\ref{sec:bias-meas} below.
The bias of our sample is then compared to the bias forecast for the BGS sample. In this first iteration, the two values will not match. Therefore, we adjust the initial guess $\sigma_{L_\mathrm{min}}$ to obtain a better agreement. Here we use the fact that increasing $\sigma_{L_\mathrm{min}}$ tends to decrease the bias, as more low-mass objects are selected in exchange for fewer high-mass ones. The adjustment is repeated until we obtain an acceptable agreement between the bias of our sample and $b_\mathrm{BGS}$.

\begin{table}
\caption{
Properties of our synthetic galaxy catalogue. $\mathrm{log}L_\mathrm{min}$ and $\sigma_{L_\mathrm{min}}$ are the parameters used in the probability function, Eq.~\eqref{eq:hod-prob}, to select objects that reproduce the 
DESI BGS specifics. $N_\mathrm{sources}$ and $b$ are the number of objects and the galaxy bias in each redshift bin, respectively.
}
\begin{center}
\adjustbox{max width=\columnwidth}{
\begin{tabular}{cc c c c}
 \toprule
$\bar{z}$ & $\mathrm{log}L_\mathrm{min}$ & $\sigma_{L_\mathrm{min}}$\Bstrut  & $N_\mathrm{sources}$ & $b$ \\
 \hline
 0.25  & $10.04$ &   $10^{-5}$\Tstrut & 7'875'200 & $1.290 \pm 0.012 $  \\
 0.35  & $10.4$ &   $0.21$    & 3'020'151  & $1.605 \pm 0.016 $ \\
 0.45  & $11.13$  &  $0.59$    & 495'698 &  $1.70 \pm 0.02 $ \\
\bottomrule
\end{tabular}}
\end{center}
\label{tab:hod-sel}
\end{table}

In Table~\ref{tab:hod-sel} we report the values of $\mathrm{log}L_\mathrm{min}$ and $\sigma_{L_\mathrm{min}}$
used in the final selection, together with the number of sources $N_\mathrm{sources}$ and the
galaxy bias of our synthetic galaxy catalogue $b$. 
In the redshift bin centred at $\bar{z} = 0.25$, despite the fact that we selected the most luminous objects present in the complete catalogue ($\sigma_{L_\mathrm{min}} \rightarrow 0$), the bias of our mock is lower compared to the forecast value for DESI. This is probably due to the fact that we do not include satellite galaxies in our construction --- we only consider objects located at the center of halos or subhalos.

In Fig.~\ref{fig:flux-limit-desi} we show the 
distribution of the observed fluxes for the objects selected
in each of the redshift bins. 
The fluxes are in units of $L_\odot/\mathrm{Mpc^2}$, where $L_\odot$ is the solar luminosity.
DESI will observe, as primary sample, galaxies in the $r$-band with magnitude limit $r_\mathrm{lim} = 19.5$,
see \citet{DESI:2016fyo,Ruiz-Macias:2020wxa}. 
In order to compare the fluxes of our sources to the DESI BGS limit, we convert this
magnitude limit into a flux limit. We report in Appendix~\ref{ap:flux-conv} the conversion between $r-$band magnitudes, that will be measured for the DESI BGS, and fluxes in $L_\odot/\mathrm{Mpc^2}$. Using Eq.~(\ref{eq:flux-convert}) we find that the magnitude limit $r_\mathrm{min} = 19.5$ correspond to a flux limit $F_\mathrm{lim} \approx 136\,L_\odot/\mathrm{Mpc^2}$. In Fig.~\ref{fig:flux-limit-desi} we show with a dashed black line this limiting flux. We see that all the objects included in our synthetic catalogue are above this flux limit and, therefore, should be detectable by DESI. In this case, the magnification bias of the full
sample $s(\bar{z}, F_\mathrm{cut})$ is identically zero by definition. In reality the situation is probably complicated, as the target selection of the DESI BGS will be based on criteria that may depend directly or indirectly on observed fluxes. Computing the effective magnification bias in such a situation is challenging. However, in order to have better control on the selection, we can impose an additional flux limit towards the faint end of the source population. Such a procedure can, of course, likewise be applied to the actual data, always at the expense of losing a certain fraction of objects.
In our analysis we will therefore consider two catalogues: 1) a baseline catalogue which is not flux limited and includes all sources listed in Table~\ref{tab:hod-sel}, and 2) a flux-limited catalogue, obtained from the baseline catalogue applying a redshift-dependent magnitude cut.

\begin{figure}
\begin{center}
\adjustbox{max width=\columnwidth}{
  \includegraphics[width=0.6\textwidth]
{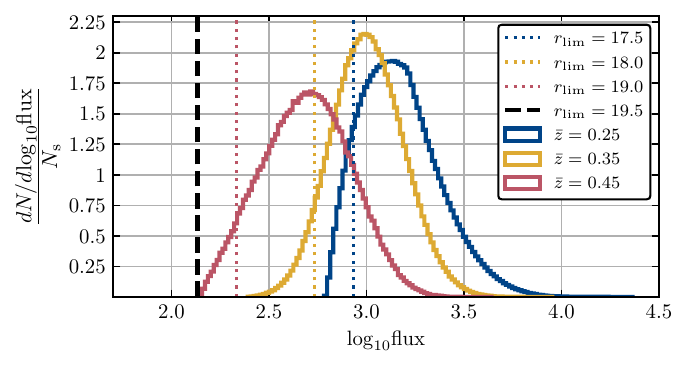}
}
  \end{center}
  \caption{Normalised distribution of fluxes in the three redshift bins considered in our analysis. A vertical black dashed line highlight the DESI flux limit, corresponding to a magnitude in the r-band of $r_\mathrm{min} = 19.5$. The colored dotted lines highlight the minimum values of the fluxes used to construct the flux-limited catalog. }
  \label{fig:flux-limit-desi}
\end{figure}

In Table~\ref{tab:flux-lim-catalogue} we report the magnitude/flux limit applied to the baseline catalogue to obtain the flux-limited catalogue,
together with the number of objects found in each redshift bin, and their galaxy bias. 
The flux-limited catalogue contains fewer and brighter objects, thus it has slightly lower number density than the DESI Bright Galaxy Survey, and slightly higher galaxy bias compared to our baseline mock. 
The selection of flux limit values in our study was based on two criteria. We do not want to remove too many sources, to limit the impact on the number density of our catalogue, but at the same time we want a flux limit well inside the distribution, to have a non-negligible magnification bias $s(z, F_{\rm lim})$. 

\section{Separation of sources into a bright and faint population}
\label{sec:split}

In order to measure the relativistic dipole, we need
to separate the objects in this catalogue into two samples that we identify
as `bright' and `faint' populations, respectively. The reference quantity to perform this split
is the observed flux. 
Since the flux is inversely proportional to the square of the luminosity distance, sources with the same intrinsic luminosity are typically brighter 
at low redshift. In order to ensure that the two populations have the same
redshift distribution, we follow a procedure similar to \citet{Alam:2017izi}, and we choose a redshift-dependent flux cut for each redshift bin.  
In a first step, we divide the redshift bins in $20$ sub-bins. In
each of the sub-bins, we estimate the flux cut $F^i_\mathrm{cut}$ as the $N$th percentile of the fluxes for all sources found within the sub-bin $i$. 
The equal split correspond to the percentile $N = 50$, i.e. the median flux is the threshold that separates bright and faint in each sub-bin. 
The flux cut decreases monotonically with the mean redshift of each sub-bin $z_i$, and we can interpolate to obtain a smooth function of the flux cut $F_\mathrm{cut}(z)$. We use this smooth function to separate our sample in bright and faint: for each object in the sample with observed redshift $z_k$,  we estimate the $F_\mathrm{cut}(z_k)$. If the object has flux below $F_\mathrm{cut}(z_k)$, it is classified as `faint', otherwise it is considered as `bright'. This smooth splitting ensures us that we have the same number of bright and faint galaxies at each redshift, which is essential in order not to introduce a spurious systematic dipole.

While an equal split between bright and faint objects minimizes the shot noise contribution to the covariance, it
does not necessarily translate into the highest signal-to-noise, due to the fact that the dipole signal depends on a non-trivial combination of the bias and magnification bias parameters of the two populations. We will therefore explore different flux cuts for our measurements.  
In the remainder of this section, we outline the procedure to estimate the galaxy bias, the magnification bias, and the evolution bias from our catalogues (Sec.~\ref{sec:bias-meas}), 
and we show the dependence of the biases on the splitting strategy (Sec.~\ref{sec:bias-diff}). More details on the estimation of the biases, and the associated errors, are given in Appendix~\ref{ap:bias-meas}.

\begin{table}
\caption{
Properties of our flux-limited mock catalogue. $r_\mathrm{lim}$ and $F_\mathrm{lim}[L_\odot/\mathrm{Mpc^2}] $
are the magnitude/flux cut applied in each redshift bin to the baseline mock to obtain this sample.  
$N_\mathrm{sources}$ and $b$ are the number of objects and the galaxy bias in each redshift bin, respectively.
}
\begin{center}
\adjustbox{max width=\columnwidth}{
\begin{tabular}{cc c c c c}
 \toprule
$\bar{z}$ & $r_\mathrm{lim}$ & $F_\mathrm{lim}[L_\odot/\mathrm{Mpc^2}] $\Tstrut\Bstrut  & $N_\mathrm{sources}$ & $b$ & $s$\\
 \midrule
 0.25  & $17.5$ &   $858.13$    & 7'007'463\Tstrut & $1.31 \pm 0.01 $  & $0.24$  \\
 0.35  & $18$ &     $541.44$    & 2'820'017  & $1.63 \pm 0.02 $ & $0.11$ \\
 0.45  & $19$  &    $215.55$    & 466'977 &  $1.74 \pm 0.02 $ & $0.10$     \\
\bottomrule
\end{tabular}}
\end{center}
\label{tab:flux-lim-catalogue}
\end{table}

\subsection{Measurement of galaxy bias, magnification bias and evolution bias}
\label{sec:bias-meas}

The theory prediction of the dipole depends on the galaxy bias, the magnification bias, and the evolution bias of the bright and faint population. 
All these quantities can be estimated independently of the dipole measurements.  

The galaxy bias for all the populations (bright, faint, and full) is measured following the method described in \citet{Lepori:2022hke}. We construct a full-sky map of the number counts using the observed angular position of the sources, we estimate the angular power spectrum of this map using the \texttt{HEALPix} library \citep{Gorski:2004by}, and we fit the angular power spectrum on large scales to a theory model that has the linear bias as free parameter. In the fit we include multipoles in the range $\ell \in [20, 80]$.  
On the theory side, RSD are modelled using the Kaiser prescription which is 
a good approximation for the wide redshift bins considered in our analysis, see e.g.~\cite{Lepori:2022hke}. 
Note that this procedure can be applied to synthetic catalogues since we know the cosmology of the simulation. 
In a real surveys, this procedure would not work since the cosmology is unknown. The correct procedure in this case is to leave the biases as free parameters, and to combine measurements of the dipole with measurements of the monopole, quadrupole and hexadecapole for each of the populations. This allows to constrain the biases of the two populations together with the dipole amplitude. Here we do not do that, since our goal is to demonstrate that the dipole is detectable in a sample like the BGS, and that the measurements agree with the theoretical prediction. To this end we use the biases measured from the angular power spectra as input for our theoretical predictions.

The magnification bias is estimated in slightly different ways for the flux-limited catalogue and the bright/faint populations. 
For the flux-limited catalogue, we estimate the magnification bias in each redshift bin, $s(z, F_\mathrm{lim})$, from the cumulative luminosity/flux function of the baseline catalogue. 
Following Eq.~\eqref{eq:sbias-def}, we compute its logarithmic derivative at the flux limits reported in Table~\ref{tab:flux-lim-catalogue}. In Fig.~\ref{fig:sbias-extimate-fcut} we show the distribution of fluxes for the objects in the redshift bin $\bar{z} = 0.25$ (yellow histogram), and the corresponding cumulative distribution (blue line). A dashed black line highlights the flux limit applied to construct the flux-limited catalogue. The measured values of the magnification bias $s$ are reported in Table~\ref{tab:flux-lim-catalogue}. 
For the bright population, the procedure is slightly different since we have a redshift-dependent flux cut separating the bright and faint populations. Therefore, we construct a redshift-dependent magnification bias for the bright objects in $20$ sub-bins
following the same procedure outlined for the flux-limited full catalogue, and estimate an effective value of the magnification bias
by taking the average of these values, weighted by the number of bright objects in the sub-bins (see Appendix~\ref{ap:bias-meas} for more detail). 
Finally, the magnification bias of the faint population, as discussed in Sec.~\ref{sec:biases-th}, is related to the magnification bias of the bright and full catalogues. Therefore, we simply compute it from Eq.~\eqref{eq:sfaint}. 
Note that this procedure can also be applied to real data. Any uncertainty in the measurements of the magnification biases would translate directly into an uncertainty on the theoretical predictions. We account for this uncertainty (computed in Appendix~\ref{ap:bias-meas}) when estimating the significance of the detection.

\begin{figure}
\begin{center}
\adjustbox{max width=\columnwidth}{
  \includegraphics[width=0.6\textwidth]
{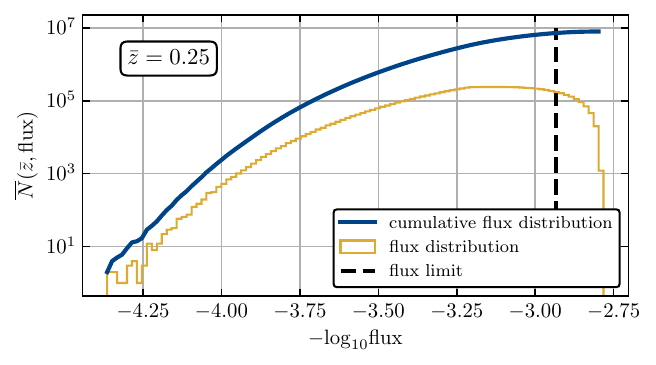}
}
  \end{center}
  \caption{Distribution of fluxes in the baseline catalogue, at $\bar{z} = 0.25$. 
  The dashed vertical line indicates the flux limit we apply in order to obtain the flux-limited catalogue. 
  Its magnification bias is estimated from the slope of the cumulative flux distribution (blue curve), see Eq.~\eqref{eq:sbias-def}. 
  }
  \label{fig:sbias-extimate-fcut}
\end{figure}

The evolution bias of our sample can be estimated from Eq.~\eqref{eq:fevo-2} or Eq.~\eqref{eq:fevo-3}, depending on how the selection is done. In our case, we have chosen the flux cut to separate the bright and faint populations such that we have an equal number of bright and faint galaxies in each thin redshift sub-bin. As a consequence, the flux cut evolves with redshift, but the corresponding intrinsic luminosity cut is constant (as the redshift dependence of the flux accounts for the luminosity distance dependence). We can therefore use Eq.~\eqref{eq:fevo-2} for $\bar{L}_{\rm cut}(z)=\bar{L}_{\rm cut}={\rm constant}$ to compute the evolution bias. We see immediately that in this case the evolution biases of the two populations are the same. As a consequence,
Eq.~\eqref{eq:dip} contains only one term that depends on the evolution bias multiplied by the bias difference of the two tracers (the first term in the third line). Since the bias differences are significantly smaller than the magnification bias differences, the contribution of the evolution bias to the dipole signal is negligible. For this reason, we can safely set the values of evolution bias to the fiducial values $\fe_{\rm B}=\fe_{\rm F} = 0$.
For the flux-limited catalogue the situation is slightly different. By artificially imposing a flux limit, we remove from our catalogues the faintest objects, preferentially found at high redshift. As a result, the redshift distribution of the objects inside the bin is not very realistic (see Fig.~\ref{fig:dNdz-test}). This unrealistic distribution has a direct impact on the evolution bias, and to model it properly we need to model the constant flux limit within the redshift bins. In Appendix~\ref{ap:evo-bias}, we propose a modelling of the evolution bias in this case and we show that it can produce a minor systematic effect on the predicted signal. However, this effect is not significant enough to impact the outcome of our analysis. Therefore, in the rest of the analysis, we also set the evolution bias to zero in the flux-limited catalogues. The estimated values of galaxy and magnification bias are reported in Table~\ref{tab:spec}.

\subsection{Dependence of biases on the split strategy}
\label{sec:bias-diff}
As shown in Eq.~\eqref{eq:dip}, for flux-separated bright and faint catalogues, the amplitude of the dipole signal depends not only on the bias difference,
$b_{\rm B} - b_{\rm F}$, but also on the magnification bias difference, $s_{\rm B} - s_{\rm F}$, and the combination, $b_{\rm F} s_{\rm B} -  b_{\rm B} s_{\rm F}$. 
\begin{figure}
\begin{center}
\adjustbox{max width=\columnwidth}{
  \includegraphics[width=0.6\textwidth]
{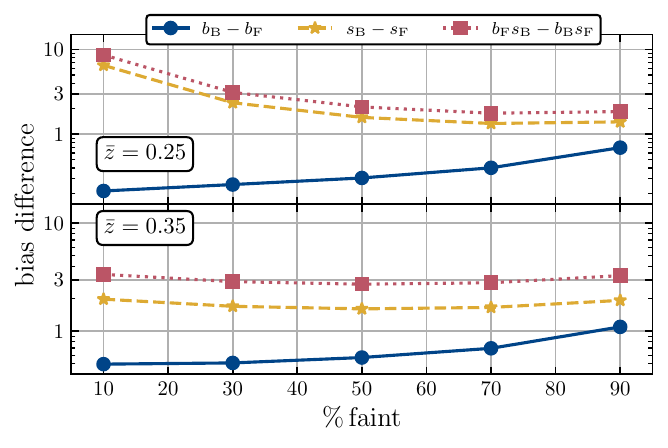}
}
  \end{center}
  \caption{Bias differences as function of the percentage of faint sources for the redshift bins centered at $\bar{z} = 0.25$ (top panel) and  $\bar{z} = 0.35$ (bottom panel). }
  \label{fig:biases-percfaint}
\end{figure}
In Fig.~\ref{fig:biases-percfaint} we show these three quantities for the two redshift bins centred at
$\bar{z} = 0.25, 0.35$, for five different splits of bright and faint galaxies (10\%, 30\%, 50\%, 70\% and 90\% of faint galaxies).

The difference in galaxy bias ranges between $0.2$ and $0.7$ at $\bar{z} = 0.25$, and between $0.5$ and $1.1$ at $\bar{z} = 0.35$. In both redshift bins it is maximised choosing a large percentage of faint objects, to be cross-correlated to fewer bright and highly biased objects. 

The differences in magnification biases has a different behaviour, however. From Fig.~\ref{fig:biases-percfaint} we see that at $\bar{z}=0.25$ the terms related to magnification biases are maximised for a small percentage of faint objects. At $\bar{z}=0.35$ the terms are roughly independent on the choice of splitting. This behaviour is due to the fact that two competing effects are at play. On the one hand, reducing the percentage of faint galaxies implies moving the flux cut toward the faint end of the luminosity function. The cumulative luminosity function is less steep at the faint end, and this translates into a smaller value of the magnification bias of the bright population $s_\B$. On the other hand, the magnification bias of the faint population (in the case with no flux limit) is simply given by $s_\F = -s_\B \bar{N}_\B/\bar{N}_\F$. Decreasing the fraction of faint objects therefore boosts the absolute value of $s_\F$ by a factor $\bar{N}_\B/\bar{N}_\F$, consequently increasing the magnification bias difference. 

At $\bar{z} = 0.25$ the second effect seems to dominate, leading to a difference, $s_{\rm B} - s_{\rm F}$, that is significantly larger for catalogues with a small percentage of faint objects. 
At $\bar{z} = 0.35$, on the other hand, the two effects seems to roughly compensate, and we see that the magnification bias difference depends only mildly
on the fraction of faint and bright objects. 
We also note that the difference in magnification bias is significantly larger than the difference in galaxy bias for flux-selected populations, reaching $s_{\rm B} - s_{\rm F}\sim 6.5$ and $s_{\rm B} - s_{\rm F} \sim 2$ at $\bar{z} = 0.25$ and $\bar{z} = 0.35$, respectively. 
For this reason, the mixed product difference, $b_{\rm F} s_{\rm B} -  b_{\rm B} s_{\rm F}$, exhibits a similar trend as 
the magnification bias difference.

To summarise, from Fig.~\ref{fig:biases-percfaint} we expect that at $\bar{z}=0.25$ the amplitude of the dipole is maximised for a large fraction of bright galaxies, while at $\bar{z}=0.35$ we expect a roughly constant amplitude for all splits. Obviously, the detectability of the dipole depends not only on the amplitude of the signal, but also on the covariance, which increases for unequal splitting (due to shot noise). In the following, we therefore illustrate two ways of partitioning: a) an equal split, with roughly $50\%$ bright objects and $50\%$ faint objects (the configuration that minimises the shot-noise covariance), and b) an unequal split, with roughly $90\%$ bright objects and $10\%$ faint objects (a configuration that maximizes the magnification bias difference and therefore the signal). We perform the analysis in both configurations for the baseline catalogue and for the flux-limited catalogue. In Table~\ref{tab:spec} we report the specifics for these four settings considered in our work. Note that in the bin $\bar{z} = 0.45$, the number of sources is too small to obtain reliable bias measurements for case b). Therefore, at this redshift the equal split is chosen as default.

\begin{table}
\caption{Estimated specifics for the four cases considered in our analysis. For the full catalogue, and the flux-selected bright and
faint populations, we report the number of objects on the full sky, $N_\mathrm{sources}$, their effective redshift $\bar{z}_\mathrm{eff}$, the estimated values of galaxy and magnification bias and the number of objects used in the measurement of the relativistic cross-correlation dipole over a BGS-sized sub-region of the full sky, $N_{\mathrm{sub}}$ (see Sec.~\ref{sec:measurement} for details). Details on the estimation of effective redshift and biases can be found in Sec.~\ref{sec:bias-meas} }
\label{tab:spec}
\begin{center}
\adjustbox{max width=\columnwidth}{
\begin{tabular}{ c c c c c c }
 \toprule
 \multicolumn{6}{c}{ {\bf Case 1: $50\%$ faint,  $50\%$ bright no flux limit} } \\
 \midrule
  \multicolumn{6}{c}{$\bar{z} = 0.25$} \\
Type & $N_\mathrm{sources}$ & $\bar{z}_\mathrm{eff}$ & $b$ & $s$ & $N_{\mathrm{sub}}$\\
FULL   &  7'875'200  & $0.2560$ &  $1.290 \pm 0.012 $ & $0$ & 2'647'734\\
FAINT  &  3'939'224  & $0.2560$ & $1.148 \pm 0.011 $ & $-0.791\pm 0.009$ & 1'323'440 \\
BRIGHT &  3'935'976   &  $0.2560$& $1.437 \pm 0.014 $ & $0.791\pm 0.009$ & 1'324'294\\
 \midrule
   \multicolumn{6}{c}{$\bar{z} = 0.35$} \\
Type & $N_\mathrm{sources}$ & $\bar{z}_\mathrm{eff}$ & $b$ & $s$ & $N_{\mathrm{sub}}$\\
FULL   &  3'020'151   & $0.3534$  & $1.605 \pm 0.016 $ & $0$ & 1'014'681    \\
FAINT  &  1'510'340   & $0.3534$  & $1.356 \pm 0.015 $ & $-0.81\pm 0.02 $ & 508'116\\
BRIGHT &  1'509'811   &  $0.3534$ & $1.866 \pm 0.019 $ & $0.81\pm 0.02 $  & 506'565\\
 \midrule
    \multicolumn{6}{c}{$\bar{z} = 0.45$} \\
Type & $N_\mathrm{sources}$ & $\bar{z}_\mathrm{eff}$ & $b$ & $s$ & $N_{\mathrm{sub}}$\\
FULL   &  495'698   & $0.4525$  & $1.70 \pm 0.02 $ & $0$ & 166'173 \\
FAINT  &  247'788   & $0.4525$  & $1.32 \pm 0.03 $ & $-0.58 \pm 0.03$ & 83'049\\
BRIGHT &  247'910   &  $0.4525$ & $2.08 \pm 0.03 $ & $0.58\pm 0.03$ & 83'124 \\
 \bottomrule\noalign{\vspace{6pt}}
 \toprule
 \multicolumn{6}{c}{ {\bf Case 2: $10\%$ faint,  $90\%$ bright no flux limit} } \\
 \midrule
  \multicolumn{6}{c}{$\bar{z} = 0.25$} \\
Type & $N_\mathrm{sources}$ & $\bar{z}_\mathrm{eff}$ & $b$ & $s$ & $N_{\mathrm{sub}}$\\
FULL   &  7'875'200  & $0.2560$ &  $1.290 \pm 0.012 $ & $0$  & 2'647'734\\
FAINT  &  790'551  & $0.2560$ & $1.110 \pm 0.013 $ & $-5.84\pm 0.21$ & 264'415\\
BRIGHT &  7'084'649   &  $0.2560$& $1.314 \pm 0.013 $ & $0.65\pm 0.02 $ & 2'383'319 \\
 \midrule
   \multicolumn{6}{c}{$\bar{z} = 0.35$} \\
Type & $N_\mathrm{sources}$ & $\bar{z}_\mathrm{eff}$ & $b$ & $s$ & $N_{\mathrm{sub}}$\\
FULL   &  3'020'151   & $0.3534$  & $1.605 \pm 0.016 $ & $0$  & 1'014'681 \\
FAINT  &  301'973     & $0.3534$  & $1.18 \pm 0.02 $ & $-1.79\pm0.07 $ & 101'345\\
BRIGHT &  2'718'178   &  $0.3534$ & $1.654 \pm 0.017 $ & $0.198\pm0.008$ & 913'336 \\
 \bottomrule\noalign{\vspace{6pt}}
  \toprule
 \multicolumn{6}{c}{ {\bf Case 3: $50\%$ faint,  $50\%$ bright  with flux limit}} \\
 \midrule
  \multicolumn{6}{c}{$\bar{z} = 0.25$} \\
Type & $N_\mathrm{sources}$ & $\bar{z}_\mathrm{eff}$ & $b$ & $s$ & $N_{\mathrm{sub}}$\\
 FULL  &  7'007'463   & $0.2520$ &  $1.31 \pm 0.01 $ & $0.24$  & 2'356'046\\
FAINT  &  3'507'107   & $0.2520$ & $1.16 \pm 0.01 $ & $-0.34 \pm 0.05 $ & 1'178'444\\
BRIGHT &  3'500'356   &  $0.2520$& $1.45 \pm 0.01 $ & $0.82 \pm 0.05$ & 1'177'602 \\
 \midrule
  \multicolumn{6}{c}{$\bar{z} = 0.35$} \\
 Type & $N_\mathrm{sources}$ & $\bar{z}_\mathrm{eff}$ & $b$ & $s$ & $N_{\mathrm{sub}}$\\
FULL   &  2'820'017   & $0.3516$  & $1.63 \pm 0.02 $ & $0.11$ & 947'845 \\
FAINT  &  1'410'719   & $0.3516$  & $1.380 \pm 0.015 $ & $-0.65\pm 0.06 $ & 474'971 \\
BRIGHT &  1'409'298   &  $0.3516$ & $1.89 \pm 0.02 $ & $0.88\pm 0.06 $ & 472'874   \\
 \midrule
    \multicolumn{6}{c}{$\bar{z} = 0.45$} \\
 Type & $N_\mathrm{sources}$ & $\bar{z}_\mathrm{eff}$ & $b$ & $s$ & $N_{\mathrm{sub}}$\\
FULL   &  466'977   & $0.4511$  & $1.74 \pm 0.02 $ & $0.10$ & 156'469 \\
FAINT  &  233'485   & $0.4511$  & $1.38 \pm 0.03 $ & $-0.43 \pm 0.04 $ & 78'120\\
BRIGHT &  233'492   & $0.4511$  & $2.10 \pm 0.03 $ & $0.63\pm 0.04   $ & 78'349 \\
 \bottomrule\noalign{\vspace{6pt}}
  \toprule
 \multicolumn{6}{c}{ {\bf  Case 4: $10\%$ faint,  $90\%$ bright with flux limit}} \\
 \midrule
  \multicolumn{6}{c}{$\bar{z} = 0.25$} \\
Type & $N_\mathrm{sources}$ & $\bar{z}_\mathrm{eff}$ & $b$ & $s$ & $N_{\mathrm{sub}}$ \\
 FULL  &  7'007'463   & $0.2520$ &  $1.31 \pm 0.01 $ & $0.24$ & 2'356'046 \\
FAINT  &  706'124     & $0.2520$ & $1.10 \pm 0.01 $ & $-3.6\pm 0.5$ & 236'839\\
BRIGHT &  6'301'339   &  $0.2520$& $1.33 \pm 0.01 $ & $0.67\pm 0.05 $ & 2'119'207 \\
 \midrule
   \multicolumn{6}{c}{$\bar{z} = 0.35$} \\
Type & $N_\mathrm{sources}$ & $\bar{z}_\mathrm{eff}$ & $b$ & $s$ & $N_{\mathrm{sub}}$\\
FULL   &  2'820'017   & $0.3516$  & $1.63 \pm 0.02 $ & $0.11$ & 947'845 \\
FAINT  &  282'157     & $0.3516$  & $1.22 \pm 0.02 $   & $-1.57 \pm 0.8 $ & 95'387 \\
BRIGHT &  2'537'860  &  $0.3516$ & $1.68 \pm 0.02 $ & $0.30\pm 0.09$ & 852'458 \\
 \bottomrule
\end{tabular} 
}
\end{center}
\end{table}

\section{Measurements}
\label{sec:measurement}

Here we present the measurement of the dipole for the four cases described above. In order to mimic the DESI BGS
we perform this measurement on a sub-region of the full sky defined by $\cos \theta > 1/3$, where $\theta$ is the zenith angle, which corresponds to an area of 13'820 square degrees. The resulting number of galaxies used in the measurement for each case is listed in the last column of Table~\ref{tab:spec}. The BGS actually has a more complicated geometry but this should not significantly affect the signals on the scales where we perform the measurements.

In Sec.~\ref{sec:dipole}, we describe the estimator for the signal, in Sec.~\ref{sec:covmeas} we discuss the measurements of the covariance and in Sec.~\ref{sec:covpred} we compute the uncertainty in the theoretical signal. The results are discussed in Sec.~\ref{sec:results}, and finally in Sec.~\ref{sec:significance} we compute the significance of the dipole detection.

\subsection{Dipole estimator}
\label{sec:dipole}

We estimate the full two-point cross-correlation function of our bright and faint data catalogues with a modified version of the publicly avaible code CUTE \citep{Alonso:2012}, which employs the Landy-Szalay estimator \citep{Landy:1993yu}:
\begin{equation}
\label{eq:LS}
    \xi_\textrm{BF}(d,\mu) = \frac{D_\mathrm{B}D_\mathrm{F}-D_\mathrm{B}R_\mathrm{F} -R_\mathrm{B}D_\mathrm{F}+R_\mathrm{B}R_\mathrm{F}}{R_\mathrm{B}R_\mathrm{F}}\,.
\end{equation}
The original version of CUTE assumes symmetry along the line of sight since it was designed to measure the even multipoles of the correlation function. Here we follow~\citet{Breton:2018wzk} to generalise the implementation of the Landy-Szalay estimator in a way that allows for an asymmetry in the cross-correlation along the line of sight.

In Eq. (\ref{eq:LS}), $d$ is the separation between a pair of galaxies 
and $\mu$ is the cosine of the angle between the line of sight to the respective pair and their separation vector. We choose the line of sight as the one connecting the observer with the midpoint of the separation vector, in order to minimize wide-angle effects~\citep{Gaztanaga:2015jrs}. The estimator uses data catalogues $D_i$ and corresponding random catalogues $R_i$ which follow the same redshift distribution as $D_i$, but have a uniform distribution per solid angle. The combinations of the $D_i$ and $R_i$ in Eq. (\ref{eq:LS}) correspond to histograms of pairs between the respective catalogues, binned in $d$ and $\mu$ and normalised by the total number of possible pairs between both catalogues. In this work we bin $d$ in $20$ bins with $d\leq 120\,\Mpc/h$ and $\mu$ in $500$ bins with $-1<\mu<1$. We use random catalogues that are ten times larger than the data catalogues. Since the catalogues consists of angular patches of approximately 14'000 square degrees, with no holes nor irregular boundaries, there is no need to build denser random catalogues. 

From the full two-point cross-correlation function,  the multipoles can be extracted via the relation
\begin{equation}
    \xi_\ell(d) = \frac{2\ell+1}{2} \int_{-1}^1 \xi_\mathrm{BF}(d,\mu) P_\ell(\mu) d\mu\,,
\end{equation}
where $P_\ell(\mu)$ is the $\ell$-th Legendre polynomial. Here, we approximate this integral with a sum over the bins in $\mu$ to get the dipole via
\begin{equation}
    \xi_\textrm{1}(d) \simeq \frac{3}{2}\sum_{\mu=-1}^{\mu=1} \xi_\textrm{BF}(d,\mu) \mu \Delta \mu\,,
    \label{eq:estimator}
\end{equation}
where $\Delta\mu$ is the width of the $\mu$-bins.

Note that in this paper we only consider cross-correlations of two populations of galaxies. In this case, as shown in~\cite{Bonvin:2015kuc}, the optimal estimator is given by Eq.~\eqref{eq:estimator}. On the other hand, if one splits galaxies into more than two populations, the estimator can be optimised by weighting the cross-correlations by the corresponding bias difference. 

\subsection{Estimating the covariance}
\label{sec:covmeas}

We estimate the covariance using the jackknife method \citep{Norberg:2008tg}:
\begin{align}
 {\rm cov}^{\rm JK}_{ij}&\equiv {\rm cov}^{\rm JK}(\xi_1(d_i),\xi_1(d_j))\label{eq:covJK}\\
 &= \frac{(N-1)}{N}\sum^N_{k=1}\left[\xi^k_1(d_i) - \bar{\xi}_1(d_i)\right]\left[\xi^k_1(d_j) - \bar{\xi}_1(d_j)\right]\, ,\nonumber
\end{align}
where we split the survey volume into $N=100$ sub-volumes of approximately equal area and shape by running a {\sc kmeans}\footnote{\url{https://github.com/esheldon/kmeans_radec/}} algorithm \citep{DES:2016qvw, DES:2015eop} on a random catalogue with uniform angular distribution. We then apply the same split to the data catalogues and we measure the dipole $N$ times, by removing each time a different sub-volume $k$ from the survey. $\xi^k_1(d_i)$ denotes the measurement when the sub-volume $k$ has been removed and  $\bar{\xi}_1(d_i)$ is the mean over the $N$ measurements:
\begin{equation}
\bar{\xi}_1(d_i) = \sum^N_{k=1} \frac{\xi^k_1(d_i)}{N}\, .
\end{equation}
The number of sub-volumes has been chosen such that they encompass the largest separation $d_{\rm max} = 120$\,Mpc$/h$ at $\bar{z} = 0.25$. 
We have furthermore checked that by increasing the number of sub-volumes from 32 to 100, the covariance changes by less than 20\%. Further increasing the number of regions to 150 has little impact on the covariance, which indicates that using 100 jackknife regions provides a robust estimate of the covariance.

It is important to mention that jackknife covariances contain a non-negligible amount of noise, in particular when considering the off-diagonal terms.
However, the inverse of a noisy covariance matrix is in general a biased estimate. Therefore, to correct for such a bias we take into account the so-called Hartlap factor. Although it is not mathematically exact in the case of non-independent realisations, it was shown that it 
still provides accurate results in this scenario~\citep{Hartlap2007}. In summary, when we require the inverse of the jackknife covariance matrix, for example to obtain the detection significance (see Sect.\,\ref{sec:significance}), we multiply the inverse covariance matrix by $(N_{\rm JK}-p-2)/(N_{\rm JK}-1)$, where $N_{\rm JK}$ is the number of jackknife regions (100 in our case) and $p$ stands for the number of data points (in our case there are 16 bins for separations above 27\,Mpc$/h$).

Comparing the jackknife covariance with the theory covariance, ${\rm cov}_{ij}^{\rm th}$ given in Eq.~\eqref{eq:cov_theo}, we find that the theory covariance is smaller 
by around 25\,\% for the range of separations used here.
Calculating the significance using the theory prediction may therefore overestimate the detectability of the dipole. On the other hand, the non-diagonal terms in the jackknife covariance are very noisy, which may introduce spurious effects when the covariance is inverted.
In order to capture any contribution to the covariance that is not included in our theoretical model, while keeping the matrix easily invertible, we consider a third option to determine the final covariance. This consists in combining the theory and jackknife covariances. Intuitively, we keep the shape of the theory covariance (i.e. the correlation matrix, corr$^{\rm th}$), but rescale it by the diagonal of the jackknife covariance. The same approach was used for example in~\citet{Prat,Sanchez,Crocce}. In more detail, we compute the combined matrix as:
\begin{equation}
\label{eq:cov_comb}
    \text{cov}^{\rm comb}_{ij}=\text{corr}^{\rm th}_{ij}\sqrt{\text{cov}^{\rm JK}_{ii}\text{cov}^{\rm JK}_{jj}}\,,
\end{equation}
where $i$ and $j$ run over the two axes of the covariance.

\begin{figure}
    \centering
    \includegraphics[width=0.47\textwidth]{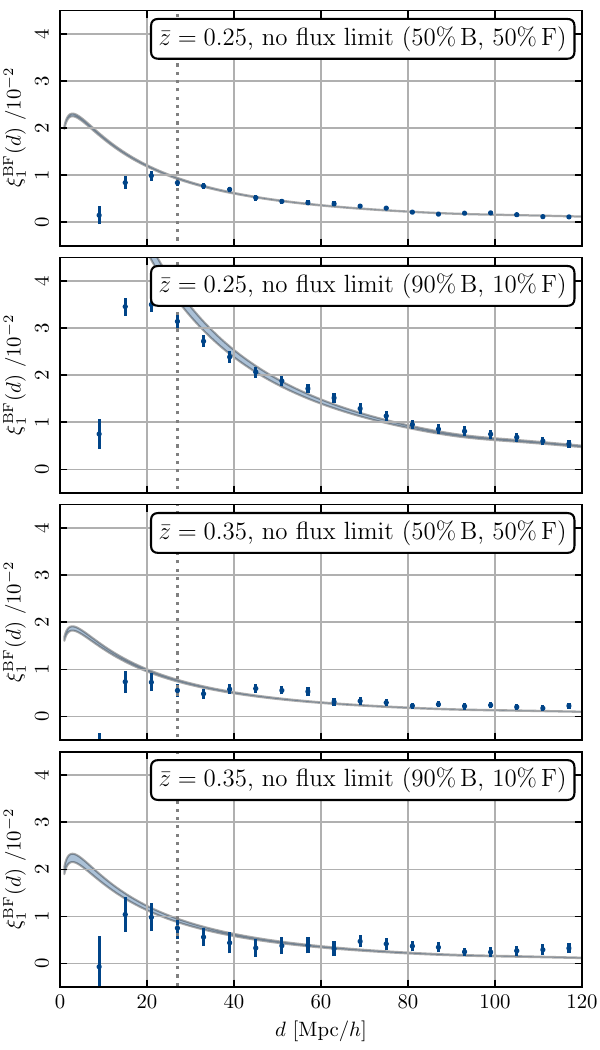}
    \caption{The measured 2PCF dipoles of the catalogues without flux limit at redshifts $\bar{z}=0.25$ and $\bar{z}=0.35$ for the two different splitting ratios (50\% bright and 50\% faint, 90\% bright and 10\% faint) compared to the prediction from linear theory. The shaded region covers the uncertainty of the theoretical prediction due to errors in the bias measurements. The grey vertical dotted line marks the bin with $d = 27\,\mathrm{Mpc}/h$. The error bars are derived from the jackknife estimate described in Sec.~\ref{sec:covmeas}.}
    \label{fig:dipole_no-flux-limit}
\end{figure}

\begin{figure}
    \centering
    \includegraphics[width=0.47\textwidth]{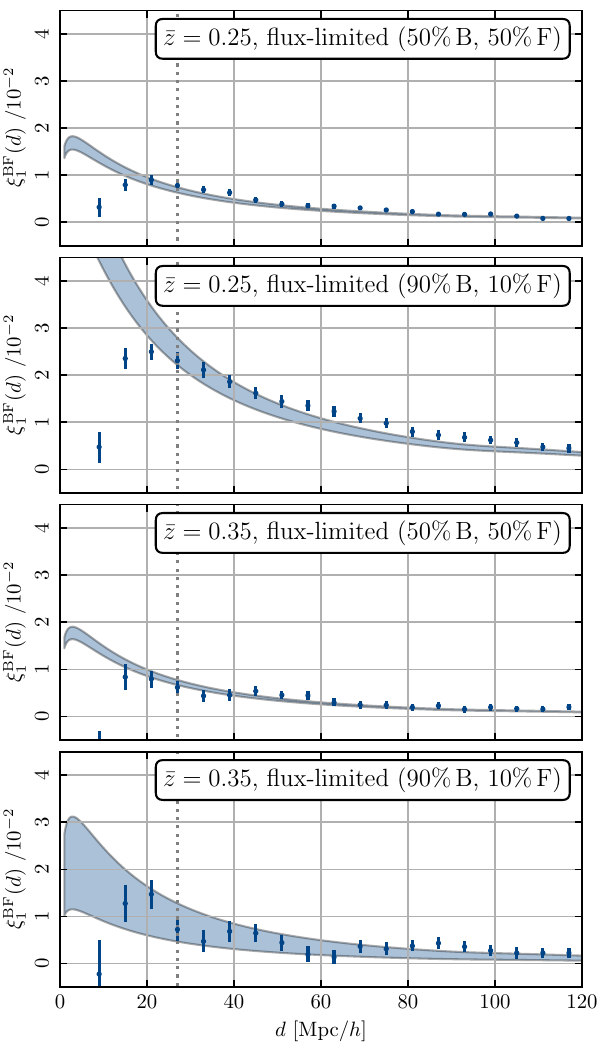}
    \caption{The measured 2PCF dipoles of the catalogues with flux limit at redshifts $\bar{z}=0.25$ and $\bar{z}=0.35$ for the two different splitting ratios (50\% bright and 50\% faint, 90\% bright and 10\% faint) compared to the prediction from linear theory. The shaded region covers the uncertainty of the theoretical prediction due to errors in the bias measurements. The grey vertical dotted line marks the bin with $d = 27\,\mathrm{Mpc}/h$. The error bars are derived from the jackknife estimate described in Sec.~\ref{sec:covmeas}.}
    \label{fig:dipole_flux-limit}
\end{figure}

\begin{figure}
    \centering
    \includegraphics[width=0.47\textwidth]{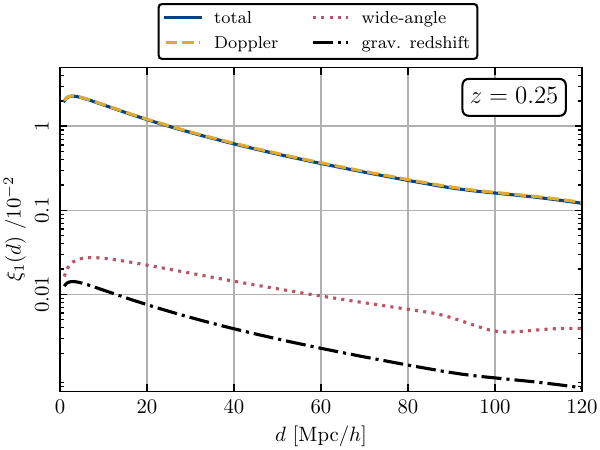}
    \caption{The different contributions to the dipole at $\bar{z}=0.25$ using the bias and magnification bias measured from the bright and faint catalogues. The yellow dashed line contains the Doppler terms, the pink dotted line the wide-angle effects (that are negative), the black dot-dashed line the gravitational redshift and the blue line is the total.}
    \label{fig:dipole contributions}
\end{figure}

\subsection{Uncertainty on the theoretical prediction}
\label{sec:covpred}

In addition to the uncertainty on the measurement there is an uncertainty on the theoretical prediction due to the uncertainty in the determination of the galaxy biases and magnification biases, given in Table~\ref{tab:spec}. The uncertainty on the bias of the faint and bright populations are uncorrelated. These uncertainties are also uncorrelated with the magnification bias uncertainties. However, from Eq.~\eqref{eq:sfaint} we see that the magnification biases of the bright and of the faint population are not independent. More precisely, the uncertainty on $s_\F$ is due to the uncertainty on $s_\B$ and the uncertainty on $s(z,F_{\rm lim})$, i.e.\ the magnification bias of the full sample. For the catalogues without flux limit, this second contribution vanishes. For the catalogues with flux limit, this contribution is there, but we expect the uncertainty on $s(z,F_{\rm lim})$ to be completely negligible with respect to that on $s_\B$. $s(z,F_{\rm lim})$ can indeed be measured from the cumulative number of galaxies in the whole population, in the redshift bin in question, with a fixed flux cut over the entire bin. As explained in Sec.~\ref{sec:bias-meas}, $s_\B$ is instead measured from the bright population, in thin redshift sub-bins, containing a much smaller number of galaxies. As such, even for the catalogues with flux limit, we neglect the error from $s(z,F_{\rm lim})$.

The covariance related to the theoretical prediction can then be written as
\begin{align}
\label{eq:covpred_deriv}
{\rm cov}^{\rm pred}_{ij}=& \sum_{\substack{\alpha,\beta= \\ b_\B,b_\F,s_\B}}\frac{\partial\xi_1(d_i)}{\partial\alpha}{\rm cov}(\alpha,\beta)\frac{\partial\xi_1(d_i)}{\partial\beta}\, .
\end{align}
Denoting the variance of each term by 
\begin{align}
 {\rm cov}(b_\B,b_\B)=\Delta b_\B^2\,  ,\  {\rm cov}(b_\F,b_\F)=\Delta b_\F^2\,, \  {\rm cov}(s_\B,s_\B)=\Delta s_\B^2   
\end{align}
and using that the uncertainties on $b_\B, b_\F$ and $s_\B$ are uncorrelated we obtain
\begin{equation}
\label{eq:covpred}
{\rm cov}^{\rm pred}_{ij}= \sigma_{b_\B}\!(d_i)\, \sigma_{b_\B}\!(d_j)+\sigma_{b_\F}\!(d_i)\, \sigma_{b_\F}\!(d_j)
+\sigma_{s_\B}\!(d_i)\, \sigma_{s_\B}\!(d_j)\,,
\end{equation}
where
\begin{multline}
\sigma_{b_\B}\!(d_i)=\Bigg\{\frac{\HH}{\HH_0}\left[f\left(\frac{2}{r\HH}+\frac{\dot{\HH}}{\HH^2} \right)+5s_\F f\left(1-\frac{1}{r\HH} \right)\right]\nu_1(d_i)\\
-\frac{2}{5}f \frac{d_i}{r}\mu_2(d_i)\Bigg\}\Delta b_\B\, ,
\end{multline}
\begin{multline}
\sigma_{b_\F}\!(d_i)=\Bigg\{-\frac{\HH}{\HH_0}\left[f\left(\frac{2}{r\HH}+\frac{\dot{\HH}}{\HH^2} \right)+5s_\B f\left(1-\frac{1}{r\HH} \right)\right]\nu_1(d_i)\\
+\frac{2}{5}f \frac{d_i}{r}\mu_2(d_i)\Bigg\}\Delta b_\B\, ,
\end{multline}
\begin{multline}
\sigma_{s_\B}\!(d_i)=\frac{\HH}{\HH_0}\Bigg\{-3\left(1+\frac{\bar{N}_\B}{\bar{N}_\F}\right)f^2\left(1-\frac{1}{r\HH} \right)\\
-5\left(b_\B\frac{\bar{N}_\B}{\bar{N}_\F}+b_\F \right) f\left(1-\frac{1}{r\HH} \right)
\Bigg\}\nu_1(d_i)\Delta s_\B\, .
\end{multline}

We note that the uncertainty in the bias measurements from actual observations may differ to some degree from the uncertainties presented in this work. As stated earlier, in this work we determine the linear biases at each redshift by comparing the angular power spectra of galaxies with that of the dark matter. In a real survey, the biases would be determined directly from the even multipoles. 

Concerning the magnification bias: the procedure to determine it in real data will be similar to the one used here. For simplicity, in this work the magnification bias has been determined  from the full-sky catalogues, but it could of course be determined from the reduced samples, as will be done with real data. This may slightly change the uncertainties and consequently the detection significances for the BGS compared to what is reported here.

\subsection{Results}
\label{sec:results}

Fig.~\ref{fig:dipole_no-flux-limit} shows the dipole at $\bar{z}=0.25$ and $\bar{z}=0.35$ for the two catalogues without flux limit (50\% of bright and 90\% of bright). The corresponding results with flux limit are shown in Fig.~\ref{fig:dipole_flux-limit}. The error bars on the measurement are obtained from the jackknife covariance matrix, Eq.~\eqref{eq:covJK}, while the shaded regions represent the theoretical uncertainty from Eq.~\eqref{eq:covpred}.

In the cases without flux limit, the theory prediction agrees very well with the measurement, down to $\sim$27\,Mpc/$h$, confirming the validity of linear perturbation theory above this scale. At smaller scales, the linear prediction is significantly larger than the measured dipole, which exhibits a turnover around 20\,Mpc/$h$. A nonlinear modelling is therefore required to describe those scales. When calculating the significance of the detection we use only scales above 27\,Mpc/$h$. In Fig.~\ref{fig:dipole contributions} we show the different effects contributing to the relativistic dipole at $\bar{z}=0.25$ for the case with equal splitting and no flux limit. We see that the dipole is dominated by Doppler effects, that are boosted by the factor $1/(r\HH)$ at low redshift compared to the gravitational redshift contribution. Note that a way of increasing the gravitational redshift contribution would be to use density-split clustering~\citep{Paillas:2022wob} to build the two populations of galaxies, rather than flux split. Such splits would allow to choose populations with large bias differences and consequently boost the amplitude of gravitational redshift~\citep{Beutler:2020evf}.

Comparing the different catalogues, we see that as expected from Fig.~\ref{fig:biases-percfaint}, at $\bar{z}=0.25$ the amplitude of the dipole is larger for the case with 90\% of bright galaxies. At $\bar{z}=0.35$, on the other hand, there is no significant boost for the 90\% case, which is also consistent with Fig.~\ref{fig:biases-percfaint}. The errors are slightly larger for the 90\% case, due to a larger shot noise. We also see that the dipole decreases significantly with redshift due to the $1/(r\HH)$ coefficient in Eq.~\eqref{eq:dip}. The errors, on the other hand, increase with redshift. This is due to the fact that errors are dominated by shot noise (see Fig.~\ref{fig:covariance}) which quickly increase with redshift due to lower number densities of galaxies. From this we expect the significance of the measurement to be even larger in the two lowest redshift bins of the BGS sample: $\bar{z}=0.05$ and $\bar{z}=0.15$. In the higher redshift bin, $\bar{z}=0.45$, we find that the dipole is too small to be robustly detected, see Figs.~\ref{fig:dipole_0.45_nolim} and~\ref{fig:dipole_0.45_lim} in Appendix~\ref{ap:dipole_0.45}.

Fig.~\ref{fig:dipole_flux-limit} shows the results in the case with a flux limit. We see that overall the flux limit reduces the amplitude of the dipole, which is however still very robustly detected. This is not surprising, since adding a flux limit makes the magnification bias of the faint sample less negative, see Eq.~\eqref{eq:sfaint}, which in turn reduces the difference in the magnification bias between the two populations. Since our flux limit was chosen somewhat arbitrarily, the importance of this effect may be somewhat different in the true BGS sample. What we see however is that adding a flux limit does not alter the detectability of the dipole at low redshift significantly. 

The errors are also slightly larger in the case with flux limit, which again is due to the increase of shot noise since we have removed a fraction of galaxies by imposing a flux limit. 
We also see that the theoretical uncertainties are larger in the case with flux limit. 
By removing mainly sources at high redshift we generate a somewhat artificial distribution in redshift (see Fig.~\ref{fig:dNdz-test} in Appendix~\ref{ap:evo-bias}). As a consequence the magnification bias fluctuates significantly within the redshift bin. Since we estimate the error on the magnification bias from these fluctuations (see Eq.~\eqref{eq:errorsB}), it is necessarily larger. Note that this effect would probably be smaller in a real flux-limited sample that is expected to have a smoother redshift evolution.

Finally, we see that at $\bar{z}=0.25$, for the case with 90\% of bright galaxies, the agreement between theory and measurement is slightly worse than in the other cases at scales between 50\,Mpc/$h$ and 100\,Mpc/$h$. This may be due to the evolution bias that we neglect in the theoretical prediction. Whereas for the case without flux limit the evolution bias is small and contributes in a negligible way to the signal, in the case with flux limit it may not be true anymore due to the artificial distribution of sources with redshift shown in Fig.~\ref{fig:dNdz-test}. In Appendix~\ref{ap:evo-bias} we show a measurement of $f^{\rm evo}$, taking into account the redshift distribution of our sample, and we find that including this value in the theoretical prediction leads to a very good agreement with the measured dipole. Again, the measured value of $f^{\rm evo}$ should not be taken too seriously, since it is related to the specific distribution of our sample, but it is a demonstration that even in the case where $f^{\rm evo}$ is not negligible, it can be measured from the sample and included in the modelling.

\subsection{Significance of the dipole detection}
\label{sec:significance}

In this last section before concluding we are interested in estimating the significance of the detection of the relativistic dipole in the synthetic galaxy catalogues used in this analysis. We determine the detection in terms of $\sigma$ by the difference in $\chi^2$ between considering or rejecting the presence of a dipole. In more detail, let us consider the $\chi^2$ when we account for a relativistic dipole in our prediction:
\begin{equation}
    \chi^2_{\rm dipole} = \sum_{ij} \left(\xi_{1,i}^{\rm meas}-\xi_{1,i}^{\rm pred}\right)^{\rm T}\left(\text{cov}^{-1}\right)_{ij}\left(\xi_{1,j}^{\rm meas}-\xi_{1,j}^{\rm pred}\right)\,,
    \label{eq:chi2with}
\end{equation}
where \text{cov} stands for the covariance of the dipole, and $i,j$ run over the different separation bins. We choose as minimum separation $27$\,Mpc/$h$ since below that our linear modelling is not valid, and as maximum separation 117\,Mpc/$h$ since adding larger scales does not increase the significance any further. Let us also consider the $\chi^2$ when no dipole is present in the prediction (null hypothesis):
\begin{equation}
    \chi^2_{\rm no\,dipole} = \sum_{ij} \left(\xi_{1,i}^{\rm meas}-0\right)^{\rm T}\left(\text{cov}^{-1}\right)_{ij}\left(\xi_{1,j}^{\rm meas}-0\right)\,.
    \label{eq:chi2without}
\end{equation}
We can then compute the detection in units of $\sigma$ as:
\begin{equation}
    \text{detection}[\sigma]= \sqrt{\Delta \chi^2}=\sqrt{\chi^2_{\rm no\,dipole}-\chi^2_{\rm dipole}}\,.
    \label{eq:detection}
\end{equation}

The dipole covariance in Eqs.~\eqref{eq:chi2with} and~\eqref{eq:chi2without} is the sum of the prediction covariance ${\rm cov}^{\rm pred}$ and the covariance on the measurement. Both do indeed affect the significance of the detection and they are independent. For the prediction covariance we use Eq.~\eqref{eq:covpred}, while for the measurement covariance we consider three cases: the theory covariance~\eqref{eq:cov_theo}, the jackknife covariance~\eqref{eq:covJK},  and the combined covariance~\eqref{eq:cov_comb}. The results for all cases are shown in Table\,\ref{tab:detection}.

In case 1 (50\% of bright sources with no flux limit), we obtain the highest detection significance at the lowest redshift, being 15.37\,$\sigma$, 14.23\,$\sigma$, or 14.42\,$\sigma$ using a theory, jackknife, or combined covariance, respectively. We note that the level of significance does not depend much on the covariance used.\footnote{Because of this we will only quote the result obtained with a combined covariance in the rest of the discussion. All results can be found in Table\,\ref{tab:detection}.} 
However, we obtain a slightly higher value with the theory covariance compared to the jackknife one. This is expected given the smaller error bars of the former. When combining the covariances, we obtain a significance in between the theory and jackknife cases. When moving to higher redshift we can observe a decrease on the detection significance of roughly a factor of two, leading to 7.58\,$\sigma$ for the combined covariance. We note that a decrease in the detection is expected, given that the largest amplitude of the dipole is expected at low redshift and that the shot noise also increases with redshift. The detection for $\bar{z}=0.35$ is however still very significant. On the other hand, at $\bar{z}=0.45$ we have not been able to compute the detection significance and therefore these cases appear with a dash in Table\,\ref{tab:detection}. The main reason for this is the decrease in the amplitude of the dipole and the decrease of the number of sources. Both effects will make the observations more compatible with a null signal. In this specific case, the difference of $\chi^2$ values present in Eq.\,(\ref{eq:detection}) becomes negative. 
Such a scenario can occur when the data points are very much compatible with a null signal. Just because of the statistical noise, a null prediction might provide a slightly better $\chi^2$ than the dipole prediction.

Focusing now on the second case (90\% of bright objects with no flux limit), we see an increase in the detection significance at $\bar{z}=0.25$, reaching 18.94\,$\sigma$. As seen in Sec.~\ref{sec:results}, the signal is significantly larger in this case, while the noise only slightly increases, leading to a boost in the significance. This split is therefore better than the equal split.
At $\bar{z}=0.35$ on the other hand, the signal is only very slightly larger for 90\% of bright than for the equal split, and therefore the increase of shot noise actually leads to a decrease in the detection significance, down to 5.01\,$\sigma$.

Cases 3 and 4 correspond to cases 1 and 2 but applying a flux limit. In both cases we find that the detection is higher at low redshift, as before. At $\bar{z}=0.25$, for the equal split, the significance decreases to 9.03\,$\sigma$. This corresponds to a reduction by a factor 1.6 compared to the case without flux limit. Such a degradation is partly due to the change on the magnification bias of the faint sample by applying this additional flux limit. In addition, less sources are present in the catalogues, which increases the shot noise. And finally, as explained in Sec.~\ref{sec:results} the error on the theoretical prediction is also larger in this case. All these effects combine to reduce the significance. 
The degradation at $\bar{z}=0.35$ is a bit smaller, leading to 6.02\,$\sigma$.

In the 10\%-90\% split, we obtain a degradation of roughly a factor 2.5 compared to the case without flux limit, both at $\bar{z}=0.25$ and $\bar{z}=0.35$.  This larger degradation is mainly due to the stronger increase of the prediction covariance compared to the equal splitting. Since this effect is directly related to the arbitrary redshift distribution of sources that we obtain by imposing the flux limit, the significance may actually be less degraded in a real survey. Note that although the detection significance decreases in the case of a flux limited sample, we still obtain a very significant detection at several sigma for two different redshift bins.

\begin{table}
\caption{Significance (in units of sigma) of the detection of the relativistic dipole for the 4 cases considered in our analysis.}
\label{tab:detection}
\begin{center}
\adjustbox{max width=\columnwidth}{
\begin{tabular}{ c c c c}
 \toprule
 \multicolumn{4}{c}{ {\bf Case 1: $50\%$ faint,  $50\%$ bright no flux limit} } \\
$\bar{z}$ & cov$_{\rm th}+$cov$_{\rm pred}$ & cov$_{\rm JK}+$cov$_{\rm pred}$ & cov$_{\rm comb}+$cov$_{\rm pred}$  \\
$0.25$   &  15.37  & 14.23 &  14.42  \\
$0.35$  &  8.73  & 7.74 & 7.58  \\
$0.45$ &  --   &  -- & --  \\
 \midrule
 \multicolumn{4}{c}{ {\bf Case 2: $10\%$ faint,  $90\%$ bright no flux limit} } \\
$\bar{z}$ & cov$_{\rm th}+$cov$_{\rm pred}$ & cov$_{\rm JK}+$cov$_{\rm pred}$ & cov$_{\rm comb}+$cov$_{\rm pred}$  \\
$0.25$    &  19.61  & 16.6 &  18.94   \\
$0.35$   &  6.0  & 5.42 & 5.01 \\
 \midrule
 \multicolumn{4}{c}{ {\bf Case 3: $50\%$ faint,  $50\%$ bright  with flux limit}} \\
$\bar{z}$ & cov$_{\rm th}+$cov$_{\rm pred}$ & cov$_{\rm JK}+$cov$_{\rm pred}$ & cov$_{\rm comb}+$cov$_{\rm pred}$  \\
 $0.25$  &  9.81   & 8.22 &  9.03   \\
$0.35$  &  6.79   & 5.64 & 6.02  \\
$0.45$ &  --   &  1.08 & 0.57  \\
 \midrule
 \multicolumn{4}{c}{ {\bf  Case 4: $10\%$ faint,  $90\%$ bright with flux limit}} \\
$\bar{z}$ & cov$_{\rm th}+$cov$_{\rm pred}$ & cov$_{\rm JK}+$cov$_{\rm pred}$ & cov$_{\rm comb}+$cov$_{\rm pred}$  \\
 $0.25$  &  8.43   & 6.91 &  7.92   \\
$0.35$  &  2.12    & 1.95 & 1.97  \\
 \bottomrule
\end{tabular} 
}
\end{center}
\end{table}

\section{Conclusion}
\label{sec:conclusion}

In this paper, we show that the relativistic dipole should be detectable with the BGS sample of DESI. We have built synthetic catalogues that reproduce the characteristics of this sample at redshift $\bar{z}=0.25, 0.35$ and 0.45, and have measured the dipole in these three bins. In the highest redshift bin there is no significant detection, but in the two bins at lower redshift the significance is high, reaching 19\,$\sigma$ in the optimal case. Since the signal increases at low redshift, while the noise decreases (due to the high number density of galaxies, reducing shot noise), we expect the significance to be even larger in the lowest redshift bins of the BGS, namely $\bar{z}=0.05$ and 0.15.

We compare the measured dipole with our theoretical prediction in the linear regime, and we find a very good agreement above $\sim$~30\,Mpc/$h$, showing that in this regime, linear perturbation theory provides an accurate modelling. At smaller scales, nonlinearities become important and need to be accounted for in the modelling. A nonlinear modelling of the relativistic signal using perturbation theory has been derived in~\cite{Beutler:2020evf} for the power spectrum, showing a good agreement with simulations up to $k_{\rm max}\sim 0.4\,h/$Mpc. A more phenomenological approach has been used to model the correlation function in the nonlinear regime, which agrees well with simulations~\citep{Saga:2020tqb}, but this approach does not include the contributions from magnification bias, which turns out to be significant in the linear regime. Including nonlinear scales would be interesting since it would boost the detectability of the dipole, especially the gravitational redshift contribution \citep[see][]{Saga:2021jrh}. However, in order to be able to robustly use those scales it is necessary to have a model that includes all effects, also wide-angle effects. A general formalism of the nonlinear correlation function on the full sky, including magnification and evolution bias, has recently been derived in~\cite{Dam:2023cem}, and will be tested against simulations below 30\,Mpc/$h$.

The pipeline that we develop in this paper to measure the dipole in our synthetic catalogues and calculate the significance can be applied to data. In particular, we develop a method to split the population of galaxies into two populations (bright and faint), such that these two populations have the same redshift distribution. For this we split the original bins into thin sub-bins. In each of them we have determined the flux limit needed to reach the percentage of bright and faint galaxies that we want, and we have then interpolated this function to obtain a smooth flux cut over the whole bin. This method is necessary in order to avoid having most of the faint galaxies at high redshift and the bright galaxies at low redshift. We also use these sub-bins to measure the magnification bias of the bright population that enters our theoretical modelling.

Finally we explore various ways of splitting the galaxies. Depending on the behaviour of the magnification bias and of the galaxy bias, unequal splitting between bright and faint galaxies may be preferred. In our synthetic catalogue, we find that for a sample which has no flux limit, having a larger number of bright galaxies significantly boosts the signal in the lowest redshift bin. At higher redshift, or in the case with a flux limit, this is however not the case anymore. Since these conclusions strongly depend on the characteristics of the two galaxy populations, it will be necessary to explore different scenarios once the BGS data will be available. Let us note that here we have focused on splitting galaxies with respect to their flux, but alternative ways to perform a split in two populations may be advantageous in some cases~\citep{Beutler:2020evf}. Moreover, the detectability of the dipole could be further boosted by using more than two populations of galaxies, and optimising the estimator by appropriately weighting the different cross-correlations~\citep{Bonvin:2015kuc}.


\section*{Acknowledgements}

We thank Michel-Andr\`es Breton and Florian Beutler for useful discussions and interactions. C.\ B.\ , F.\ L.\ , S.\ S.\ and J.\ A.\ acknowledge support from the Swiss National Science Foundation. C.\ B.\ acknowledges support from the European Research Council (ERC) under the European Union's Horizon 2020 research and innovation program (grant agreement No.\ 863929; project title ``Testing the law of gravity with novel large-scale structure observables'').  Computational resources were provided by the Swiss National Supercomputing Centre (CSCS) under a pay-per-use agreement with Science IT at the University of Zurich (project ID uzh34). {P.\ F.\ acknowledges support from Ministerio de Ciencia e Innovacion, project PID2019-111317GB-C31, the European Research Executive Agency HORIZON-MSCA-2021-SE-01 Research and Innovation programme under the Marie Sk\l{}odowska-Curie grant agreement number 101086388 (LACEGAL). P.\ F.\ is also partially supported by the program Unidad de Excelencia María de Maeztu CEX2020-001058-M.}

\section*{Data availability}

Our synthetic catalogue and the data shown in our plots of the dipole measurements are available at \url{https://doi.org/10.5281/zenodo.8172950}.


\bibliographystyle{mnras}
\bibliography{dipole_desi} 


\appendix

\section{Conversion of the magnitude limit for DESI into fluxes}
\label{ap:flux-conv}

In order to compare the fluxes of our sources to the DESI BGS limit, we convert this
magnitude limit into a flux limit.
The flux cut $F_\mathrm{lim}$ can be written in terms of a flux density $f_{\nu, \mathrm{lim}}$ (in units of Jansky) as
\begin{equation}
F_\mathrm{lim}[L_\odot/\mathrm{Mpc^2}] = f_{\nu, \mathrm{lim}}[\mathrm{Jy}] \,c\, \left(\frac{1}{\lambda_\mathrm{min}} - \frac{1}{\lambda_\mathrm{max}}\right) \frac{A_\mathrm{SI} B_\odot}{C_\mathrm{\lambda}},
\end{equation}
where $c$ is the speed of light (in units of $\mathrm{m}/\mathrm{s}$), $\lambda_\mathrm{min}$ and $\lambda_\mathrm{max}$ are the minimum and maximum wavelengths in the $r$-band (in
nanometres), and  $A_\mathrm{SI}, B_\odot, C_\mathrm{\lambda}$ are conversion factors between units: $A_\mathrm{SI} = 10^{-26}\,\mathrm{W}\,\mathrm{m}^{-2}\,\mathrm{Hz}^{-1}\,\mathrm{Jy}^{-1}$ converts a flux density in Jansky into the International System of Units (SI), 
$B_\odot \approx 2.474\times 10^{18} L_{\odot}\,\mathrm{Mpc}^{-2}\,\mathrm{W}^{-1}\,\mathrm{m}^2$ converts a flux from the SI system into $L_{\odot}/\mathrm{Mpc}^2$,
and $C_\mathrm{\lambda} = 10^{-9}\,\mathrm{m}/\mathrm{nm}$ converts wavelengths from nanometres into metres. 
The flux density $f_{\nu, \mathrm{lim}}[\mathrm{Jy}]$ is related to the magnitude in the $r$-band as
\begin{equation}
r_\mathrm{min} = \frac{5}{2} \mathrm{log}_{10} \left(\frac{f_{\nu, \mathrm{lim}}[\mathrm{Jy}]}{3631 [\mathrm{Jy}]}\right).
\end{equation}
Therefore, the flux limit in units of $L_\odot/\mathrm{Mpc^2}$ can be computed as a function of $r_\mathrm{min}$ as
\begin{equation}
F_\mathrm{lim}[L_\odot/\mathrm{Mpc^2}] = 3631 \times 10^{-\frac{2}{5} r_\mathrm{min}} \,c\, \left(\frac{1}{\lambda_\mathrm{min}} - \frac{1}{\lambda_\mathrm{max}}\right) \frac{A_\mathrm{SI} B_\odot}{C_\mathrm{\lambda}}. \label{eq:flux-convert}
\end{equation}
$\lambda_\mathrm{min} = \lambda_\mathrm{eff} - \Delta\lambda/2$ and $\lambda_\mathrm{max} = \lambda_\mathrm{eff} + \Delta\lambda/2$ are computed assuming an effective wavelength midpoint $\lambda_\mathrm{eff} = 616.6\,\mathrm{nm}$ and bandwidth $\Delta\lambda = 120\,\mathrm{nm}$.

\section{Details on the measurement of galaxy bias and magnification bias}

\label{ap:bias-meas}

In this section, we report more details on the procedure we have adopted to estimate the specifics of the faint and bright catalogues, needed to evaluate the theoretical prediction and covariance for the dipole measured in our $N$-body simulations. 

The effective redshift is estimated as the average redshift over the number of objects in the full-sky catalogue, that is
\begin{equation}
\bar{z}_\mathrm{eff} = \frac{1}{N_\mathrm{sources}} \sum_i z_i\,.
\end{equation}

The galaxy bias is estimated from the angular power spectrum of the number counts of the bright/faint objects. 
We construct the pixelised map of the number counts and estimate the angular power spectrum of the map using the \texttt{HEALPix} routine \texttt{anafast}. Note that our catalogues cover the full sky, thus we do not need to correct for the effect of a mask at this stage. The angular power spectra are then fitted to a theory model that includes the contributions of density and linear RSD. The linear galaxy bias is the only free parameter in this model. 
The linear RSD model is expected to be sufficiently accurate for our bias estimate, as we are considering redshift bins with width $\sigma_z = 0.1$~\citep[see][]{Lepori:2022hke}. The accuracy degrades for thinner redshift bins, see for example \citet{Jalilvand:2019brk, Matthewson:2021rmb}. We use the \texttt{curve\_fit} routine from the \texttt{scipy} package to obtain the best-fit
linear bias and its error. The error is estimated assuming the uncertainty on the simulated data has cosmic variance and shot-noise contribution. In our fit, we include scales between $\ell_\mathrm{min} = 20$ and $\ell_\mathrm{max} = 80$. We have tested the dependence of our best-fit estimation on the range of scales used in the fit. We find that a fit performed on scales $\ell_\mathrm{min} = 100$ and $\ell_\mathrm{max} = 200$ leads to changes in the bias estimates of a few percent, which would not significantly affect the theory prediction for the dipole and its covariance.  

The magnification bias for the bright populations is estimated from the cumulative luminosity function of 
the bright population. Special care has to be taken, as the flux threshold that separates bright and faint populations is given 
by a redshift-dependent function. Therefore, we estimate the values of magnification bias in the 20 sub-bins, where the redshift-dependent flux threshold is computed for a given ratio of bright and faint source. In each sub-bin, the magnification bias is given by the slope of the cumulative flux distribution at the flux threshold. From these 20 values $s_{\rm B}^i$, we compute an effective value for the magnification bias as
\begin{equation}
s_{\rm B} = \frac{\sum N^i_{\rm B}s^i_{\rm B}}{\sum N^i_{\rm B}},
\end{equation}
where $N^i_{\rm B}$ is the number of bright objects in the sub-bin $i$, and the sum runs over the 20 sub-bins. 

The error of this estimate is computed as the square root of the variance of a weighted mean
\begin{equation}
\label{eq:errorsB}
(\Delta{s_{\rm B}})^2 = \frac{\sum N^i_{\rm B}}{\left(\sum N^i_{\rm B}\right)^2 - \sum (N^i_{\rm B})^2} \sum N^i_{\rm B} (s^i_{\rm B}-s_{\rm B})^2.
\end{equation}

The magnification bias of the faint population is computed from the values of magnification bias of the full catalogue and the bright catalogue, as in Eq.~\eqref{eq:sfaint}. Its error is estimated from the error on $s_{\rm B}$ 
using the linear propagation of the uncertainty, that is
\begin{equation}
\Delta{s_{\rm F}} = \frac{\bar{N}_\B(z)}{\bar{N}_\F(z)} \Delta{s_{\rm B}}. 
\end{equation}

\section{Impact of evolution bias on the flux-limited catalogue}
\label{ap:evo-bias}

In our analysis, we do not attempt to model the redshift distribution within the redshift bin used for dipole measurements. As a consequence, the evolution bias of our catalogues cannot provide a realistic estimate of the evolution bias for the DESI BGS sample. For this reason, we neglect this effect in our fiducial analysis. However, we can still attempt to model this effect for our catalogue.

In the case without a flux limit, estimating the evolution bias is straightforward. Our selection of bright/faint populations corresponds roughly to a sharp luminosity threshold. Therefore, we can evaluate the evolution bias from the slope of the redshift distribution of bright/faint objects, which is by construction the same for the two populations. As shown in Fig.~\ref{fig:dNdz-test}, the slope is roughly a constant within the redshift bin. We correct for a geometrical factor, as shown in Eq.~\eqref{eq:fevo-2}, and find that the evolution bias of the bright and faint populations is identical, with values ranging between $-1$ and $1$ in the redshift bins we consider. The impact of the evolution bias on the baseline catalogues is completely negligible as a result.

\begin{figure}
\begin{center}
\adjustbox{max width=\columnwidth}{
  \includegraphics[width=0.6\textwidth]
{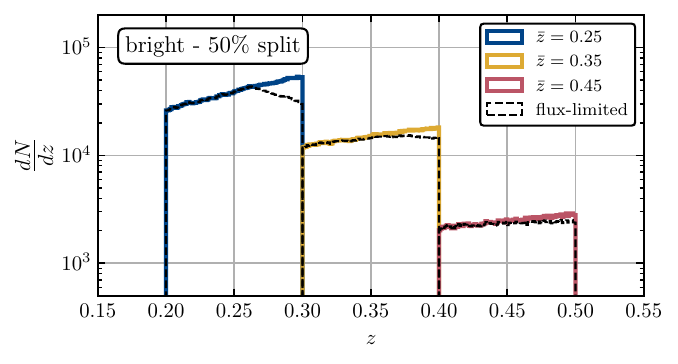}
}
  \end{center}
  \caption{Redshift distribution in the three redshift bins, for the bright populations and the equal split. Coloured histograms show the redshift distribution for our baseline case, black dashed histograms show the redshift distribution for the flux-limited catalogs.  
  }
  \label{fig:dNdz-test}
\end{figure}

Modelling the evolution bias for the flux-limited catalogue is less straightforward.
By applying a fixed flux limit in each redshift bin, we are removing the faintest objects in the bin, preferentially at high-redshift, leading to a redshift distribution that is not necessarily smooth (see dashed histograms in Fig.~\ref{fig:dNdz-test}). Furthermore, since we are not applying a sharp luminosity threshold, the evolution bias of the faint sample also depends on the slope of the luminosity function, i.e. the magnification bias, as in Eq.~\eqref{eq:fevo-3}.  
In order to take into account these two effects, we estimate the evolution bias of the bright/faint populations $f^\mathrm{evo}_{\rm B/F}(z)$ as a function of redshift inside the bin, modelling the change of slope at a transition redshift $z_{\rm t}$ whose value is computed through a fit.
The selection of the bright population is not directly affected by the flux limit. Therefore, we estimate the redshift-dependent evolution bias from Eq.~\eqref{eq:fevo-3}, that is
\begin{align}
\fe_{\rm B} (z)= - \left.\frac{\partial \mathrm{ln}  \bar{N}_{\rm B}}{\partial \mathrm{ln}(1+z)} \right|_{\rm sel}
 + \frac{\dot\HH}{\HH^2} + \frac{2}{r\HH} - 1,
\end{align}
where we denote with $ \left.\frac{\partial \mathrm{ln}  \bar{N}_{\rm B}}{\partial \mathrm{ln}(1+z)} \right|_{\rm sel}$ the logarithmic slope of the redshift distribution for the bright population. 
For the faint population, we see that the flux limit affects the redshift distribution of objects with redshift above  $z_{\rm t}$. 
Therefore, we estimate the evolution bias of the faint population,
\[
    \fe_{\rm F} (z)= \left\{\begin{array}{lr}
         - \left.\frac{\partial \mathrm{ln}  \bar{N}_{\rm F}}{\partial \mathrm{ln}(1+z)} \right|_{\rm sel}
 + \frac{\dot\HH}{\HH^2} + \frac{2}{r\HH} - 1, & \text{for } z \leq z_{\rm t}\vspace{10pt}\\
        - \left.\frac{\partial \mathrm{ln}  \bar{N}_{\rm F}}{\partial \mathrm{ln}(1+z)} \right|_{F_\mathrm{c}}
 + \frac{\dot\HH}{\HH^2} + \frac{2-5 s_{\rm lim}}{r\HH} - 5 s_{\rm lim}  - 1, & \text{for } z > z_{\rm t}
        \end{array}\right. 
\]
where we define $s_{\rm lim} \equiv s(z,F_{\rm lim})\frac{\bar{N}(z)}{\bar{N}_\F(z)}$.
From these redshift-dependent functions $\fe_{\rm B/F} (z)$, we compute an effective value for the evolution bias,
\begin{equation}
 \fe_{\rm B/F}  = \frac{\int ({\rm d}N/{\rm d}z)\,\fe_{\rm B/F} (z)\,{\rm d}z}{\int( {\rm d}N/{\rm d}z)\,{\rm d}z},
\end{equation}
where ${\rm d}N/{\rm d}z$ is the smoothed redshift distribution of the objects inside the bin.
\begin{table}
\caption{Values of the evolution bias for the flux-limited catalogue, estimated with the method described in the text.}
\label{tab:spec2}
\begin{center}
\adjustbox{max width=\columnwidth}{
\begin{tabular}{ cc c  c c}
 \toprule
 &  \multicolumn{2}{c}{ $50\%$ faint,  $50\%$ bright}  &   \multicolumn{2}{c}{ $10\%$ faint,  $90\%$ bright } \\
  \cmidrule(lr){2-3}  \cmidrule(lr){4-5}
$\bar{z}$ & BRIGHT & FAINT & BRIGHT &  FAINT\\
 \midrule
$0.25$  &  7  & 2 & 7 & -17\\
$0.35 $ &  4  &  3 & 4 & -2\\
 \bottomrule
 \noalign{\vspace{6pt}}
\end{tabular} 
}
\end{center}
\end{table}


In Table~\ref{tab:spec2} we report the values  of the evolution bias in the bins $\bar{z} = 0.25, 0.35$, where the dipole signal was detected with high statistical significance. 
We find large differences in evolution bias at $\bar{z} = 0.25$, for the  $10\%$--$90\%$ split. 

\begin{figure}
\begin{center}
\adjustbox{max width=\columnwidth}{
  \includegraphics[width=0.6\textwidth]
{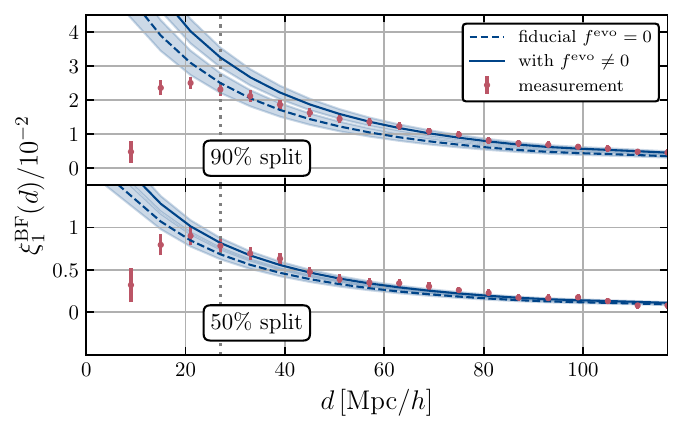}
}
  \end{center}
  \caption{Impact of evolution bias on the theory prediction for the flux-limited catalogues at $\bar{z} = 0.25$.
  Top and bottom panels refer to the $10\%$--$90\%$ and $50\%$--$50\%$ splits between faint and bright populations, respectively. 
  Dashed lines show the theory prediction used in our analysis, which sets the evolution bias to zero. Continuous lines includes the evolution bias, modelled as described in the text. 
  The measurements of the dipole are highlighted in red. 
  The shaded region represents the uncertainty of the theoretical prediction due to errors in the bias measurements (we assume here the evolution bias to be perfectly known). The grey vertical dotted line marks the scale cut used in our analysis, corresponding to the bin $d = 27\,\mathrm{Mpc}/h$.
  }
  \label{fig:fevo-test}
\end{figure}

In Fig.~\ref{fig:fevo-test} we compare the baseline theory predictions at $\bar{z} = 0.25$ (dashed lines) to a model that accounts for the values of evolution bias reported in Table~\ref{tab:spec2} (continuous lines). Including the evolution bias provides a better model for the measurements of the dipole on large scales. Therefore, the small discrepancy between theory and measurements discussed in Sec.~\ref{sec:results} for the $10\%$--$90\%$ case is most likely due to our choice of neglecting evolution bias in our modelling. Note that there remains a small disagreement at small scales, below 40\,Mpc/$h$, even after correcting for the evolution bias. Since this disagreement is not there for the $50\%$--$50\%$ case, it is probably not due to nonlinearities, but rather to some systematic effect that could affect the $90\%$--$10\%$ catalogue with flux cut. In particular, this is the catalogue with the smallest number of faint objects.

\section{Dipole measurements at \lowercase{$\bar{z} = 0.45$}}
\label{ap:dipole_0.45}
In Figs.~\ref{fig:dipole_0.45_nolim} and \ref{fig:dipole_0.45_lim} we present the measurements of the relativistic cross-correlation dipole for $\bar{z}=0.45$, respectively for the case with no flux limit and for the case where a flux limit is imposed. A theory prediction for the signal is not given for the case where the catalogue is split into 90\% bright galaxies and 10\% faint galaxies, because the measurement of the bias is not reliable due to the small number of objects.

\begin{figure}
    
    \centering
    \includegraphics[width=0.47\textwidth]{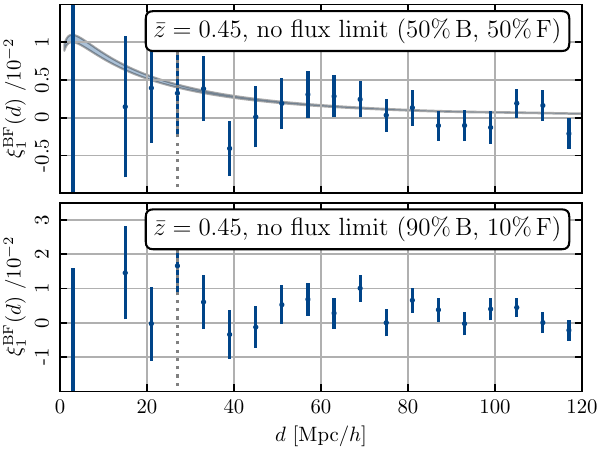}
    \caption{The measured 2PCF dipole of the $\bar{z}=0.45$ catalogue without flux limit for the two different splitting ratios (50\% bright and 50\% faint, 90\% bright and 10\% faint) compared to the prediction from linear theory. The shaded region covers the uncertainty of the theoretical prediction due to errors in the bias measurements. The grey vertical dotted line marks the bin with $d = 27\,\mathrm{Mpc}/h$. The error bars are derived from the jackknife estimate described in Sec.~\ref{sec:covmeas}.}
    \label{fig:dipole_0.45_nolim}
\end{figure}

\begin{figure}
    
    \centering
    \includegraphics[width=0.47\textwidth]{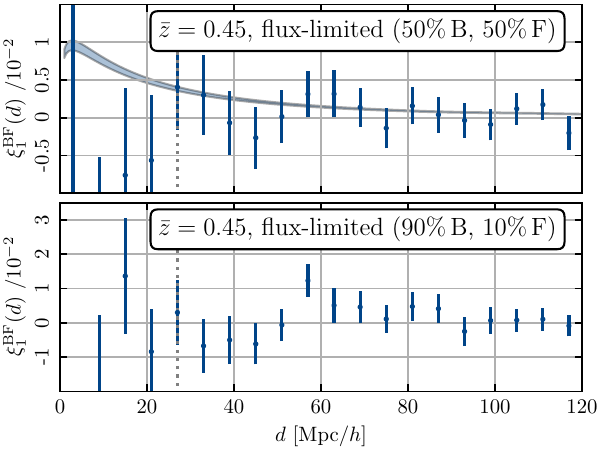}
    \caption{The measured 2PCF dipole of the $\bar{z}=0.45$ catalogue with flux limit for the two different splitting ratios (50\% bright and 50\% faint, 90\% bright and 10\% faint) compared to the prediction from linear theory. The shaded region covers the uncertainty of the theoretical prediction due to errors in the bias measurements. The grey vertical dotted line marks the bin with $d = 27\,\mathrm{Mpc}/h$. The error bars are derived from the jackknife estimate described in Sec.~\ref{sec:covmeas}.} 
    \label{fig:dipole_0.45_lim}
\end{figure}

\bsp	
\label{lastpage}
\end{document}